%% file: mn.tex
\documentclass[useAMS,usenatbib,usegraphicx]{mn2e}
\newcommand{\nuc}[1]{{\rm #1}}
\newcommand{\nucm}[2]{\ensuremath{{}^{#1}{\rm #2}}}
\newcommand{\xc}{\ensuremath{X_{\rm C}}}
\newcommand{\xn}{\ensuremath{X_{\rm N}}}
\newcommand{\xo}{\ensuremath{X_{\rm O}}}
\newcommand{\xcno}{\ensuremath{X_{\rm CNO}}}
\newcommand{\msun}{\ensuremath{M_\odot}}
\newcommand{\lsun}{\ensuremath{L_\odot}}
\newcommand{\feoh}{\textrm{[Fe/H]}}
\newcommand{\cfe}{\textrm{[C/Fe]}}
\newcommand{\nfe}{\textrm{[N/Fe]}}
\newcommand{\ofe}{\textrm{[O/Fe]}}
\newcommand{\cnratio}{\textrm{[C/N]}}

\newcommand{\he}{\rm He}
\newcommand{\xmix}{\ensuremath{X_{\rm mix}}}
\newcommand{\iip}{II$^\prime$}
\newcommand{\ivp}{IV$^\prime$}
\newcommand{\pp}{{\it p-p}}
\newcommand{\pow}[2]{\ensuremath{{#1} \times 10^{#2}}}
\newcommand{\mup}{\ensuremath{M_{\rm UP}}}
\newcommand{\pmix}{\ensuremath{P_{\rm mix}}}

\newcommand{\lows}[1]{\ensuremath_{{\rm #1}}}
\newcommand{\teff}{\ensuremath{T_{\rm eff}}}
\newcommand{\lhe}{\ensuremath{\log (L_{\rm He} / L_{\odot})}}
\newcommand{\amlt}{\ensuremath{\alpha_{{\rm MLT}}}}
\newcommand{\abra}[2]{[{\rm {#1}} / {\rm {#2}}]}

\title[Stellar Evolution of Extremely Metal-Poor Stars]{Evolution of Low- and Intermediate-Mass Stars with $\feoh \leq -2.5$}

\author[T. Suda and M. Y. Fujimoto]{T. Suda$^{1,2,3}$
\thanks{E-mail:suda@astro.keele.ac.uk}
and M. Y. Fujimoto$^{2}$\\
$^{1}$Astrophysics Group, EPSAM, Keele University, Keele, Staffordshire ST5 5BG, UK\\
$^{2}$Department of Cosmosciences, Graduate School of Science, Hokkaido University, 
Kita 10 Nishi 8, Kita-ku, Sapporo 060-0810, Japan\\
$^{3}$Research Center for the Early Universe, Graduate School of Science, 
University of Tokyo}

\begin{document}

\date{Accepted 1988 December 15. Received 2009 August 23; in original form 1988 October 11}

\pagerange{\pageref{firstpage}--\pageref{lastpage}} \pubyear{2002}

\maketitle

\label{firstpage}

\begin{abstract}
We present extensive sets of stellar models for $0.8-9.0 \msun$
in mass and $-5 \leq \feoh \leq -2$ and $Z = 0$ in metallicity.
The present work focuses on the evolutionary characteristics of hydrogen mixing into the helium-flash convective zones during the core and shell helium flashes which occurs for the models with $\feoh \la -2.5$.
Evolution is followed from the zero age main sequence to the thermally pulsating
AGB phase including the
hydrogen engulfment by the helium-flash convection during the RGB
or AGB phase.
There exist various types of mixing episodes of how the hydrogen mixing
sets in and how it affects the final abundances at the surface.
In particular, we find hydrogen ingestion events without dredge-ups that enables repeated neutron-capture nucleosynthesis in the helium flash convective zones with $\nucm{13}{C} (\alpha, n) \nucm{16}{O}$ as neutron source.
For $Z = 0$, the mixing and dredge-up processes vary with the initial mass, which results in different final abundances
in the surface.
We investigate the occurrence of these events for various initial mass and metallicity to find the metallicity dependence for the helium-flash driven deep mixing (He-FDDM) and also for the third dredge-up events.
In our models, we find He-FDDM for $M \leq 3 \msun$ for $Z = 0$ and
for $M \la 2 \msun$ for $-5 \leq \feoh \leq -3$.
On the other hand, the occurrence of the third dredge-up is limited to
the mass range of $\sim 1.5 \msun$ to $\sim 5 \msun$ for $\feoh = -3$,
which narrows with decreasing metallicity.
The paper also discusses the implications of the results of model
computations for observations.
We compared the abundance pattern of CNO abundances with observed
metal-poor stars.
The origins of most iron-deficient stars are discussed by assuming
that these stars are affected by binary mass transfer.
We also point out the existence of a blue horizontal branch
for $-4 \la \feoh \la -2.5$.
\end{abstract}


\section{Introduction}

Spectroscopic observations of extremely metal-poor (hereafter EMP) stars
in the Galactic halo with the 8m class telescopes opened a new window
to understand the formation and evolution of the Galaxy.
The statistics of these stars reveals that
(1) the frequency of stars with strong carbon enhancements is much
larger than that of Population I and II stars \citep{Rossi1999,Lucatello2006}, 
(2) the number of stars decreases significantly with decreasing metallicity
at $\feoh \la -3.5$ \citep[see e.g.,][]{Beers2005b}, and
(3) derived surface abundances of elements like carbon and
neutron-capture elements like strontium and barium show star-to-star variations
at $\feoh \la -2$ \citep{Gilroy1988,Ryan1991,McWilliam1995b,Norris1997,Aoki2002a}.

Another finding among EMP stars is the discovery of three iron-poor stars
well below $\feoh < -4$ \citep{Christlieb2002,Frebel2005,Norris2007},
which share the common feature of the strong enhancement in carbon.
It is shown that these most iron-deficient stars can be understood in terms of the evolutionary properties of $Z = 0$ or extremely metal-poor models in intermediate mass interacting binaries \citep{Suda2004,Suda2005,Komiya2007,Nishimura2009}.
Thus, the evolutionary characteristics of low- and intermediate-mass $Z = 0$
models have direct relevance to the discussions of star formation
in the early universe.

The existence of low-mass stars born with $Z = 0$ is still controversial since we have not yet determined the initial mass function (IMF) in the early epoch of the universe.
From the viewpoint of star formation theory, it is argued that low-mass stars are unlikely to form in very metal-poor clouds because of the poor sources of radiative cooling \citep[e.g.][]{Bromm2004}.
On the other hand, \citet{Lucatello2005} and \citet{Komiya2007} discussed the IMF of
EMP objects deduced from the known EMP stars and stellar evolution models.
Both results indicate the importance of the evolution of low- and intermediate-mass stars with EMP composition.
In particular, \citet{Komiya2007} propose a high-mass IMF with the medium mass of $\sim 10 \msun$ for $\feoh < -2.5$ based on the analysis of the peculiar abundance patterns of EMP stars using the models of stellar evolution at low metallicity. 
They also address the importance of the dominant roles of binaries when discussing the history of the Galactic halo stars \citep{Komiya2009a,Komiya2009b}.

One of the most prominent characteristics of models of low- and intermediate-mass EMP stars is that they have alternative channels to become carbon stars compared with metal-rich populations \citep[][hereafter FII00]{Hollowell1990,Fujimoto2000}.
For initial CNO abundances $\textrm{Z} \lows{CNO} \la \pow{3}{-7}$, the convection generated in the hydrogen-exhausted layers by the off-centre core helium flash or by the shell helium flash extends into the hydrogen-rich layers, eventually leading to the enrichment of carbon and nitrogen at the surface \citep{Fujimoto1990,Hollowell1990}.
This mechanism, called ``He-flash driven deep mixing'' (He-FDDM), is different from the third dredge-up in asymptotic giant branch (AGB) stars of Populations I and II that enriches surface material in $^{12}$C \citep{Iben1975}.

There have been many efforts in modelling the thermal pulses during AGB, and agreement is yet to be found for the resultant surface abundances after the He-FDDM.
Nevertheless, agreement is yet to be found about the resultant surface abundances after the He-FDDM, because the end products depend sensitively on the numerical treatment of mixing and nuclear burning.
Comparisons between these previous works as well as comparisons with our results will be given in later.

In this work, we describe the evolution of low and intermediate mass EMP models to explore the occurrence of a series of mixing events discovered by previous works and to revise the results of FII00 by using a stellar evolution program with updated input physics.
In particular, we intend to revise and expand a general picture of the evolution paths in the initial mass and metallicity plain proposed by FII00.
We computed a larger number of models with various initial mass and metallicity than \citet{Campbell2008b}, \citet{Lau2009}, or FII00.
The hydrogen mixing and dredge-up are also followed to obtain the surface abundances of model stars undergoing the He-FDDM events.
It is shown that different variations of mixing are obtained depending on initial mass and metallicity.
We compare our results with the observations of EMP stars.

The organisation of this paper is follows.
In Section 2, the updated version of the stellar evolution program for this work is described.
The results of the computations of stellar evolution are presented in Section 3.
Section 4 gives the discussion about the surface chemical composition of EMP stars and the implications of the comparisons between models and observations.
Conclusions follow in Section 5.

\section{The Computational Program}
The stellar evolution code used in this work is based on that of \citet{Iben1992}.
The input physics is almost the same as that in \citet{Suda2004} and \citet{Suda2007}.
We adopt \citet{Angulo1999} for nuclear reaction rates.
The conductive opacities are computed by the fitting formulae of \citet{Itoh1983}, and the neutrino loss rates by the fitting formulae of \citet{Itoh1996}.
We use the radiative opacity tables given by OPAL \citep{Iglesias1996} and by \citet{Alexander1994}.
As in \citet{Suda2007}, the interpolation of opacities between OPAL and \citet{Alexander1994} are carried out by squared sine curves for the temperature $T$ in $8000 {\rm K} < T < 10000$ K.
We construct rectangular tables to cover the regions of high temperature and high density by extrapolating with $T$ and $R \equiv \rho / T_{6}^{3}$ where $T_{6}$ is temperature in units of $10^{6}$ K.
For the extremely metal-poor models, the opacities are computed according to an interpolation recipe by Boothroyd\footnote{http://www.cita.utoronto.ca/$\sim$boothroy/kappa.html}.
However, it should be noted that we might use the wrong opacities at the surface if the CNO abundance changes greatly because \citet{Alexander1994} do not cover the opacity values for the metal abundances different from scaled solar ones.

In this work, we do not consider mass loss from the envelope, convective overshooting or semi-convection.
The borders of convective regions are determined by the Schwarzschild criterion and the mixing length is taken to be 1.5 times the local pressure scale height.

Computations are terminated after a number of thermal pulses during AGB phase or at the onset of carbon burning.
For thermally pulsating AGB (TPAGB) models, the evolutionary sequence includes the hydrogen-mixing event during the flashes and the third dredge-up events if they occur.
The hydrogen ingestion is treated by the following procedure;
the mixed protons are distributed uniformly down to a mixing depth.
The depth of the mixing is determined by the shell where the nuclear time-scale of proton capture is equal to the convective turnover time-scale.
Then, we solve the nuclear reactions in the convective zones in a time step comparable to, or shorter than the convective turnover time-scale.
The detailed numerical computations of this phenomenon require short time steps in order for the models to converge, which is time-consuming.
Because of the technical difficulty, we are not successful in computing the mixing and dredge-up events for all the models in this work.
The present work aims at clarifying the dependence of the occurrence of mixing events on initial mass and metallicity to see whether (1) hydrogen mixing takes place at the onset of the off-centre or central helium flash, (2) it occurs at the beginning of TPAGB phase, or (3) it never occurs.
We also derive the chemical abundances at the surface for light elements after the He-FDDM.

\begin{figure*}
\centering
\includegraphics[width=0.6\textwidth]{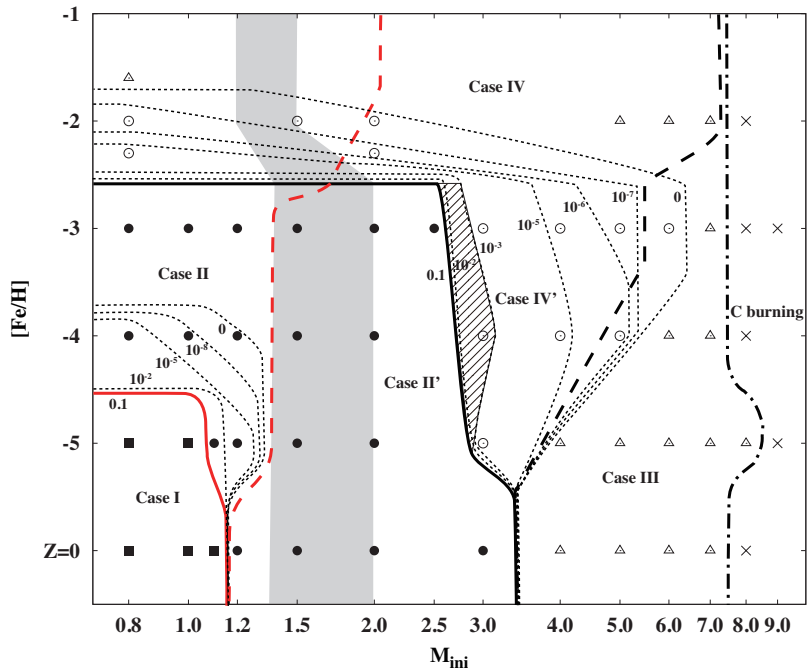}
\caption{Evolutionary characteristics of computed models on the mass-metallicity diagram, denoted by different symbols: 
		the models that undergo He-FDDM-R (filled squares), He-FDDM-A (filled circles), hydrogen ingestion event during the AGB phase but without dredge-up (open circles), AGB evolution without contact of the helium-flash convection with the hydrogen-containing layer (open triangles), and carbon burning (crosses, to the right of the dot-dashed line).
		Red and black solid lines demarcate the ranges of mass and metallicity for the He-FDDM-R and He-FDDM-A events (defined by the requirement that the helium flash convection penetrates into the hydrogen shell to $X > 0.1$), respectively.  
	    Dotted lines show the maximum extent of helium-flash convection by the hydrogen abundance of the outermost engulfed shell, attached to individual lines.
		The models with finite hydrogen abundance are regarded as undergoing hydrogen ingestion.
		The hatched area denotes the possible site for neutron-capture nucleosynthesis by the hydrogen ingestion events without He-FDDM-A.
		The red dashed line denotes the upper mass limit for off-centre ignition at the core helium flash.
		The grey shaded area shows the possible range of lower mass limit for the occurrence of the third dredge-up, which is not yet well specified.
		The black dashed line shows the upper mass limit for the occurrence of the third dredge-up.
}
\label{fig:model}
\end{figure*}

\section{Evolution of Metal-Free and Extremely Metal-Poor Models}

We compute the evolution of stars from the zero-age main sequence through TPAGB phase or through the carbon ignition for the model parameters, as shown in Figure~\ref{fig:model}, initial mass and metallicity diagram with a summary of characteristics of the resultant evolution.
Figure~\ref{fig:model} shows the mass-metallicity diagram of the computed evolutionary models and a summary of the results.
The chemical composition of initial models is the same as in FII00, i.e. $X = 0.767$, $Y = 0.233$, and $X_{3 \rm He} = \pow{2}{-5}$ in mass fraction.
For elements heavier than carbon, we adopt the scaled solar abundance of \citet{Anders1989}.
The only one exception is the model of $0.8 \msun$ and $\feoh = -1.6$ where we set $[\alpha / \textrm{Fe}] = 0.4$.
In the following subsections, we adopt the same classification of the evolution of EMP models as in FII00, which is given in Table~\ref{tab:case} and noted in Fig.~\ref{fig:model}.
With regards to the notation related to the hydrogen mixing and burning, we call the hydrogen-mixing event ``hydrogen ingestion''.
We also use the terminology ``hydrogen flash'' when the thermal runaway of burning of mixed protons occurs in the helium-flash convective zones.

\input{Table1.tex}

For Cases I or II and \iip, the model experiences He-FDDM during the core helium flash
(He-FDDM-R) or at the early phase of the TPAGB (He-FDDM-A).
The notation ``R'' and ``A'' stands for the evolutionary status at the onset of He-FDDM, i.e. the red giant branch and asymptotic giant branch phase, respectively, according to the notations in \citet{Komiya2007}.
In Fig.~\ref{fig:model}, black and red solid lines define the boundaries for the occurrence of the He-FDDM-A and the He-FDDM-R, respectively.
Dotted lines in the figure show the contours of the maximum hydrogen abundance, \xmix, (denoted by attached numerals) of the shell to which the helium-flash convection extends.
He-FDDM-R occurs only for off-centre helium core flashes, as shown by red dashed line, which denotes the boundary between the off-center and central ignition of helium.
The boundary above $\feoh = -2$ is based on the previous result with the same stellar evolution code for $\feoh = -1.6$ where we find the boundary between 2 and $2.1 \msun$ \citep{Suda2007b}.

He-FDDM-A characterizes the evolution for Cases II and \iip.
We find that the border of the occurrence of He-FDDM-A corresponds to $\xmix \ga 0.01$ where we find the hydrogen flash and the penetration of surface convection.
The difference between Case II and \iip\ is the occurrence of the third dredge-up (hereafter TDU, in Case \iip) that will bring about the enhancement of $\nucm{12}{C}$ in the envelope.
It should be noted that the boundary between these two cases is not definitive in this work because we do not find exact lower limit for the occurrence of the TDU.
Especially, for the models in the mass range near the boundary, the core mass is so small after the He-FDDM-A that it will take many pulses for the core mass to increase sufficiently for the TDU events, as discussed later, and hence, whether the TDU occurs or not will depend on the efficiency of mass loss.
The upper mass limit to the occurrence of TDU is also shown by black dashed line.
We find no or negligible amount of hydrogen ingestion for Case III, IV, and \ivp.
The absence of hydrogen flash events in Case III and \ivp~ is due to the higher entropy in the hydrogen-burning shell for more massive stars, which constitutes higher entropy barrier between the helium convective zone and the hydrogen-containing layers.
It prevents hydrogen from being engulfed by the helium-flash driven convective zones \citep{Fujimoto1990}.
This is also true for Case IV with large initial abundance of CNO elements.
The models of Cases IV and \ivp) are distinguished from the models of Case III by whether we find the TDU or not.
Some of the models of Case III may explode as carbon deflagration supernovae because they may not suffer from efficient mass loss owing to the small content of metals.
For more massive stars, they will become super AGB with electron degenerate core of oxygen, neon, and magnesium or supernovae triggered by electron capture of neon and magnesium \citep{Miyaji1980,GarciaBerro1994,Ritossa1996}.

The boundary between Cases IV and \ivp\ is defined by the different efficiency of the {\it s}-process in the \nucm{13}{C} pocket, but it is approximately taken to be the critical metallicity of $\feoh \sim -2.5$.
This is inferred from the observed properties of EMP stars which show that the abundance ratio, [Pb/Ba], does not have an increasing trend with decreasing metallicity for $\feoh \la -2.5$.
Since the ratio is expected to increase because of the decrease in the number of iron seed nuclei in the \nucm{13}{C} pockets, the observations suggest that {\it s}-process does not take place or is inefficient in the \nucm{13}{C} pocket for $\feoh \la -2.5$ as discussed in \citet{Suda2004}.

\begin{figure}
\includegraphics[width=84mm]{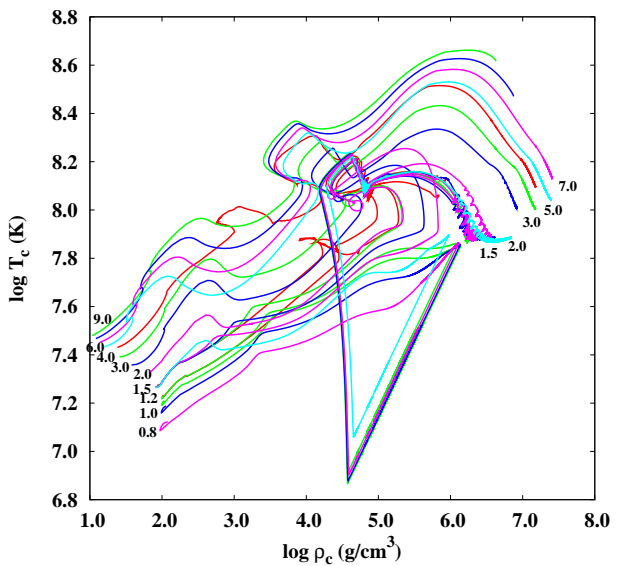}
\caption{
        Evolutionary tracks of model centres in the temperature-density 
        plane.
        Some of the models in Fig.~\ref{fig:model} are selected.
        Colour represents the metallicity of the models:
        $Z = 0$ (red), $\feoh = -5$ (green), $-4$ (blue), $-3$ (violet),
		and $-2$ (cyan).
        Labels on the panel show the model mass in units of $\msun$.
}
\label{fig:rhot}
\end{figure}

Figure~\ref{fig:rhot} shows the evolution trajectories on the central density and temperature diagram for selected models.
For $Z = 0$ stars of $M \geq 1.2 \msun$, helium burning is ignited in the centre before the neutrino losses start to play a critical role \citep{Suda2007}, while the lower mass limit of central helium ignition is $2.1 \msun$ for $\feoh = -1.3$ \citep[see, e.g.,][]{Suda2007b}.
In the present computations, the maximum initial mass of central helium ignition monotonically increases with increasing metallicity because of decrease in the temperature in the hydrogen burning shell, as found by (\cite{Cassisi1993}, see also \cite{Fujimoto1995}).
For $-5 \leq \feoh \leq -3$, in particular, the helium ignition occurs in the centre for $M \geq 1.5 \msun$.
In Fig.~\ref{fig:rhot}, the dependence of the border on the metallicity can be seen from the different evolutionary paths of helium ignition that are distinguished by the adiabatic expansion of the centre of the star, respectively.
Another variation in evolutionary paths of models at different metallicities is found in the appearance of a convective core for intermediate mass stars.
This can be recognized by the increase in the central temperature as a functio nof density due to the change of polytropic index from 3 to 1.5 during the central hydrogen burning.

Table~\ref{tab:evo} summarizes the evolutionary characteristics of our models before the He-FDDM, i.e., either of He-FDDM-R or He-FDDM-A.
The first two columns in Table~\ref{tab:evo} give the initial mass and metallicity of model.
The entries of the rest of columns differs according to whether the He-FDDM-R occurs or not.
If it occurs, the third columns gives the helium core mass just before the core helium flash, and the fifth to the eighth columns give the surface abundances of He, C, N, and O before the core helium flash, respectively.
Otherwise, the third and fourth columns give the maximum core mass interior to the hydrogen burning shell before the second dredge-up and the mass coordinate at the bottom of convective envelope after the second dredge-up, respectively.
The remaining columns give the surface abundances after the second dredge-up.

\input{Table2.tex}

\subsection{Hydrogen ingestion at the tip of RGB}

The He-FDDM-R occurs at the helium core flash if $M \la 1.1 \msun$ and $\feoh < -4$.
The border between Case I and Case II is almost consistent with that of FII00, although it has a weak metallicity dependence.
For $\feoh = -5$, we find mixing events for the models of 0.8 - $1.2 \msun$, but the models of $M = 1.1$ and $1.2 \msun$ do not result in the deep dredge-up after the hydrogen flash.
Hydrogen is engulfed only up to the shell of hydrogen abundance $\xmix \sim 10^{-2}$ and $\sim 10^{-5}$ for $1.1$ and $1.2 \msun$, respectively. 
For $\feoh = -4$, the hydrogen ingestion occurs for $M \leq 1.2 \msun$, but the hydrogen burning is not strong enough to drive convection and to dredge-up the nuclear products in the former helium convective shell.

Table~\ref{tab:hefddm} summarizes the characteristic values of He-FDDM and the surface abundances after the dredge-up.
The first and second columns specify the model parameters:
the third and fourth columns give the mass and radius of the helium-burning shell when He-FDDM takes place:
the fifth and sixth columns gives the maximum temperature reached at the bottom of helium-flash convection and the maximum helium-burning luminosity:
the next four columns provide the surface composition after the He-FDDM:
and the last column assigns the types of He-FDDM of whether it happens on the tip of the RGB (R) or during the early phase of TPAGB (A).

\input{Table3.tex}

\begin{figure}
\includegraphics[width=84mm]{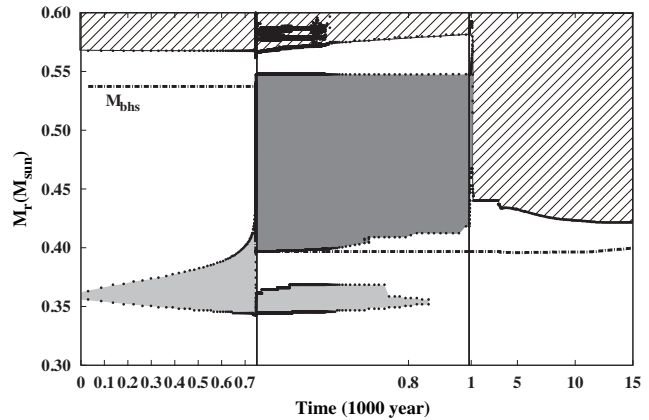}
\caption{
Time evolution of helium-flash convection due to the mixing and the dredge-up by the surface convection during the off-centre core helium flash for $0.8 \msun$ model with $\feoh = -5$.
Dash-dotted line with the label ``$M_{{\rm bhs}}$'' denotes the bottom of the hydrogen-burning shell or of the hydrogen-containing layers.
The time $t = 0$ is set at the appearance of helium-flash convective zone.
Shaded areas denote the convective zones driven by helium burning (light shaded) and hydrogen burning (dark shaded).
Hatched area represents the surface convection.
}
\label{fig:hefddmr}
\end{figure}

In the following, we give some more detailed description to the characteristics of He-FDDM-R for the model of $0.8 \msun$ with $\feoh = -5$ as a representative sincee there is no significant differences in the evolution of He-FDDM-R for the other models.
Figure~\ref{fig:hefddmr} shows the time development of convection during the off-centre core helium flash for this model.
The helium flash starts at the shell of $M_r \simeq 0.3585 \msun$ in the helium core and drives a convection there.
The helium burning rate climbs to the peak of $\log L_{\he}^{\rm max} / \lsun) = 10.86$ when the mass in the helium core is $M_{1} =  0.5395 \msun$.
  The helium-flash convective zone extends further outward and we find the hydrogen ingestion when it comes to involve the mass $0.1767 \msun$ with $\log L/\lsun = 2.976$ and $\log \teff = 3.669$.
The helium-flash convection erodes the entire hydrogen
burning shell and leaves a sharp discontinuity in the
hydrogen profile.
The mass $M_{\rm mix}$ of the hydrogen-containing layers incorporated by the helium flash-convection amounts to $\sim \pow{6}{-3} \msun$.
The mass of hydrogen engulfed is estimated at $\sim \pow{8}{-5} \msun$, and this amount is enough for the hydrogen flash to occur and split the helium-flash convective zone.
In the helium-flash convection, the mass fractions of CNO elements change before and after the hydrogen mixing from $(X_{\rm C},  X_{\rm N}, X_{\rm O} ) = (\pow{3}{-2}, \pow{2}{-9}, \pow{3}{-6})$ to $ (\pow{4.8}{-2}, \pow{9.2}{-3}, \pow{1.1}{-5})$.
The deepening of the surface convection changes the surface abundances to $\xc = \pow{1.6}{-2}$, $\xn = \pow{3.1}{-3}$, and $\xo = \pow{3.7}{-6}$.
The total mass fraction \xcno\ of CNO elements is \pow{1.91}{-2} for this model.
We do not find any hydrogen ingestion events at the AGB phase because of the large CNO abundance in the envelope after the He-FDDM-R.

Finally, we briefly comment on the weak hydrogen mixing events at the RGB so that do not involve the hydrogen flash and the splitting of convective zone.
It occurs in the models adjacent to the red line, as shown by dotted lines in Fig.~\ref{fig:model}, although the amount of hydrogen engulfed by the helium flash-convection decreases rapidly with the initial CNO abundance and with the stellar mass because of the increase in the entropy in the hydrogen burning shell \citep{Fujimoto1990}.
For example, the $1.2 \msun$ model with $\feoh = -4$ encounters a small amount ($\xmix < 10^{-8}$) of hydrogen mixing despite its large helium-burning luminosity ($\log (L_{\he}^{\rm max} / \lsun)$ = 10.57) at the peak of the core helium flash.
For very low metallicity, the dotted lines come close together since the tail of hydrogen profile is curtailed owing to the carbon production by 3$\alpha$ reactions in the bottom of hydrogen burning shell.  
   The mixed hydrogen triggers he neutron capture reactions in the helium convective zone, but in the case of He-FDDM-R, the nuclear products are buried in the core without any effect on the surface abundances.
These models simply evolve to the TPAGB and undergo He-FDDM-A.

\subsection{Hydrogen ingestion during the early phase of AGB}

The hydrogen ingestion events are found for masses up to $\sim 6 \msun$ for $\feoh = -3$, as shown by circles in Fig.~\ref{fig:model}.
In some of them, hydrogen ingestion is strong enough to cause hydrogen burning flashes which eventually drive hydrogen flash convection and deep mixing by the surface convection, giving rise to He-FDDM-A.
These Case II and \iip~models are shown by filled circles in Fig.~\ref{fig:model}.
In other models, mixed protons are burned in the helium-flash convective zones, but they do not affect the subsequent evolution except for nucleosynthesis in the helium flash convective zone.
These models are shown by open circles and we do not refer to them as He-FDDM-A.
These models are shown by open circles with the contour map of \xmix, and we do not refer to them as He-FDDM-A.

\begin{figure}
\includegraphics[width=84mm]{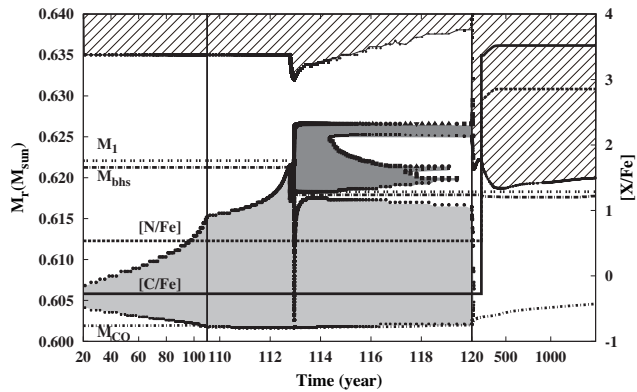}
\caption{
Time evolution of helium- and hydrogen-flash convection during the mixing and dredge-up at the fifth thermal pulse of $2 \msun$ model with $\feoh = -4$;
$M_{1}$ and $M_{{\rm bhs}}$ are the centre and bottom of the hydrogen-burning shell, defined by the shells of hydrogen abundance equal to half the surface value and equal to zero, respectively, and $M_{{\rm CO}}$ is the border between the helium-rich layer and the carbon-oxygen core, defined by the helium abundance equal to a half in mass.
The surface abundances of carbon and nitrogen are also shown in logarithms of the ratio to the scaled solar abundance relative to iron, i.e., $[ {\rm X } / {\rm Fe}]$ as indicated in the right-hands side margin.
The time $t = 0$ is set at the appearance of helium-flash convective zone.
The meanings of the shaded areas are the same as in Fig.~\ref{fig:hefddmr}.
}
\label{fig:hefddmaa}
\end{figure}

Case II and \iip\ models undergo hydrogen ingestion in the beginning of AGB after a few thermal pulses, and yet, the timing of mixing events varies with the initial mass and metallicity.
The difference depends sensitively on the thermal state of the helium core and, therefore, on the growth of the helium shell flashes.
Figure~\ref{fig:hefddmaa} shows the mixing and dredge-up by the He-FDDM-A in the fifth pulse of the $2 \msun$ and $\feoh = -4$ model as typical of the He-FDDM-A event.
In this case, we find the occurrence of weak hydrogen ingestion twice, once during the third pulse and once during the fourth pulse, neither of which causes the hydrogen flashes.
At the fifth thermal pulse, the hydrogen mixing is strong enough to drive a flash in the helium convection.
The peak luminosity of helium burning in the fifth pulse amounts to $\log (L_{{\rm He}} / L_{\sun}) = 7.28$.
Engulfed protons are mixed down to $\approx 20$ percent from the top of helium flash convection in mass, and are quickly burnt at the bottom of mixed shell where the temperature is $\sim \pow{1.0}{8}$ K.
When the helium-flash convection extends outward to the hydrogen containing shell of hydrogen abundance $X \approx 0.02$, the hydrogen-flash driven convection appears at the bottom of mixed shell.
At this moment, the hydrogen-burning luminosity reaches $\pow{1.1}{10} L_{\sun}$.
In this double convective shell phase, the outer hydrogen flash convection grows further in mass by the engulfment of the hydrogen-containing shells.
The hydrogen flash convection incorporates the mass of $\pow{8.30}{-3} \msun$ at the maximum erosion, nearly half of the maximum mass of helium-flash convection.
The hydrogen convection lasts for six years until the flash is quenched due to the expansion of the burning shell.
The luminosity due to the hydrogen burning and the temperature at the bottom of convective shell decrease to $\log (L_{{\rm H}} / L_{\sun}) = 3.55$ and $\pow{1.52}{7}$ K, respectively, just before the disappearance of hydrogen flash convection.
The abundances in the convective zone are the following; $X = 0.393$, $\xc = 0.175$, $\xn = \pow{1.37}{-2}$, and $\xo = \pow{8.22}{-3}$.

After He-FDDM-A, we have the surface abundances of $\cfe = 3.50$, $\nfe = 2.83$, and $\ofe = 1.68$ at the surface.
The mass dredged up by the surface convection is $\pow{7.89}{-3} \msun$, which is almost all the mass involved in the convective zone driven by hydrogen flash.
Since the whole envelope becomes abundant in CNO elements ([CNO/H] $> -2.5$) after the dredge-up, the hydrogen ingestion does not occur in the following thermal pulses.

\begin{figure}
\includegraphics[width=84mm]{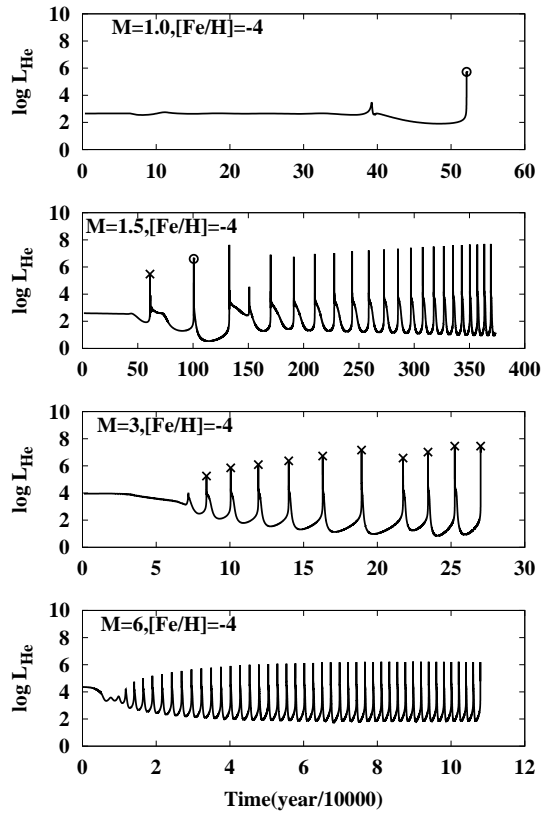}
\caption{
		Evolution of the helium-burning rate (in unit of $\lsun$) during the early TPAGB phase for different masses with the same metallicity, as given on top left corner. 
	   Crosses and circles denote the first contact of helium-flash convection with the bottom of the hydrogen-burning shell for each thermal pulse and the occurrence of He-FDDM.   
		The time $t=0$ is set at the arbitrary time in the early phase of the TPAGB.
}
\label{fig:lhemass}
\end{figure}

\begin{figure}
\includegraphics[width=84mm]{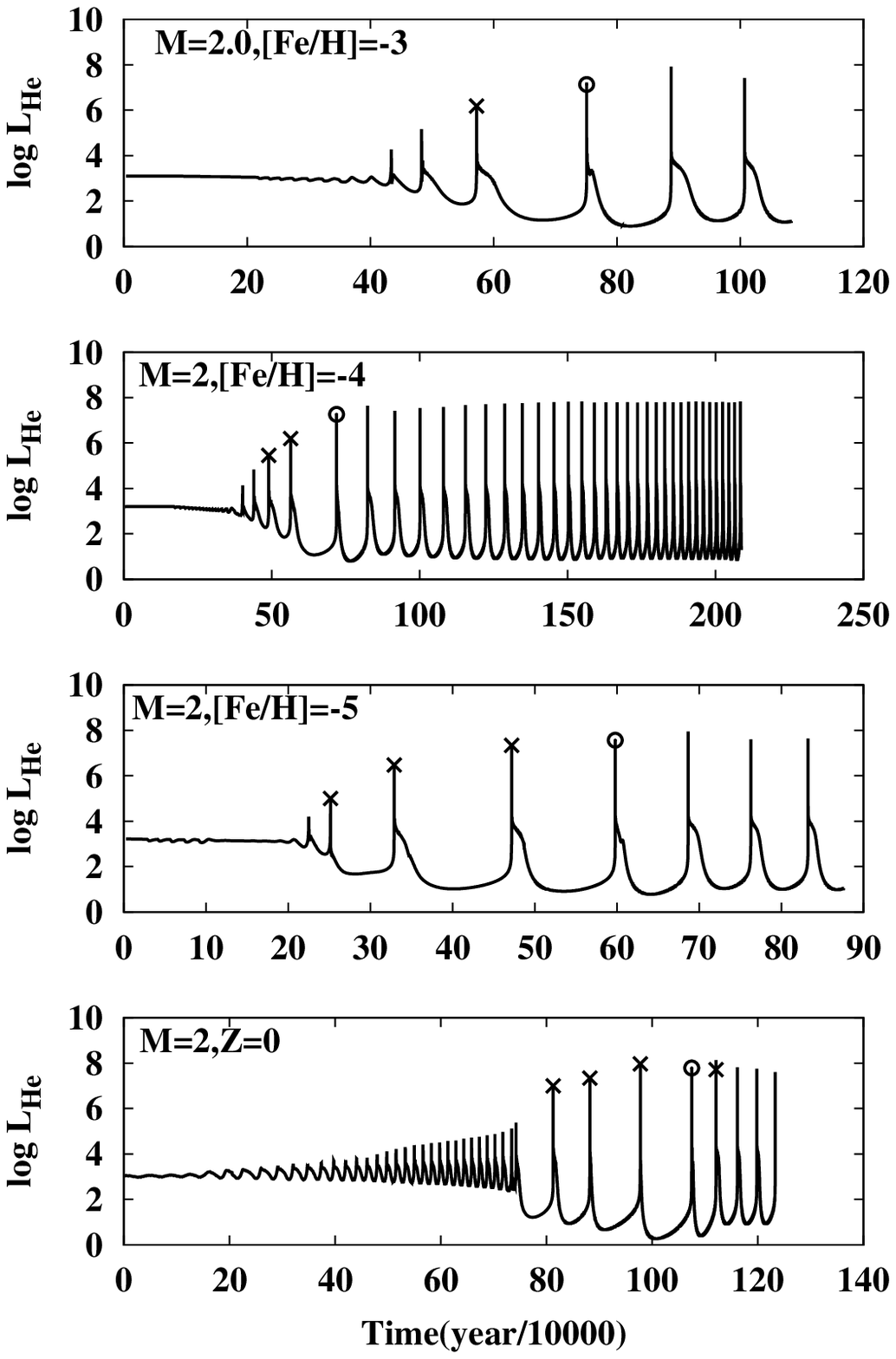}
\caption{
		The same as in Fig.~\ref{fig:lhemass}, but for different metallicity with the same mass of $M = 2 \msun$.
}
\label{fig:lhemetal}
\end{figure}

The way the thermal pulse grows differs from model to model.
Figures~\ref{fig:lhemass} and \ref{fig:lhemetal} show the variation of the growth of the thermal pulses for various models, which present the mass and metallicity sequence of shell helium flashes with a given metallicity of $\feoh = -4$ and a given mass of $M = 2 \msun$, respectively.
It shows that He-FDDM-A is preceded by weak hydrogen ingestion without hydrogen shell-flash or the splitting of convection once to several times except for the model of the smallest mass.  
The weak hydrogen ingestion may entail the neutron capture reactions with \nuc{13}C as neutron sources to produce peculiar abundance patterns of light elements, C through Al with s-process elements \citep{Nishimura2009}.
These nuclear products are incorporated into the convective zone during the subsequent helium flashes and finally are dredged up to the surface during He-FDDM-A, differently from He-FDDM-R.

In addition, we find the case of hydrogen ingestion without He-FDDM.
This is the case for the model of $3 \msun$ in Fig.~\ref{fig:lhemass}, and for the models shown by open circles in Fig.~\ref{fig:model}.
These models can be the origins of both CEMP-{\it s} and CEMP-no{\it s} depending on the efficiency of {\it s}-process nucleosynthesis in the helium-flash convective zones.
For example, TDU is found at the tenth thermal pulse for $3 \msun$ model with $\feoh = -3$ after the weak hydrogen-mixing events of $\xmix \sim 10^{-5}$.
In this model, 11 hydrogen-mixing events are found during 12 thermal pulses.

Interestingly, the final surface chemical composition after the He-FDDM-A can differ greatly not only with the mass among the models with $Z = 0$ but also between the models with $Z = 0$ and other EMP models.
For $M \la 1.5 \msun$, it depends on the strength of the thermal pulse when the hydrogen ingestion drives the flash.
For the model of $M = 1.2 \msun$ with $Z = 0$, He-FDDM-A occurs at the helium shell-flash of the maximum helium-burning luminosity, $\lhe \simeq 5.0$. 
This is relatively low compared with the typical luminosity of $\lhe \ga 6$.
As a result of a weak thermal pulse, the final surface abundances of CNO elements are as small as [C/H] $= -1.98$, [N/H] $= -0.43$, and [O/H] $= -2.66$, but sufficiently large to prohibit the hydrogen ingestion during the subsequent helium shell-flashes.
For $M = 1.5 \msun$ with $Z = 0$, we also find a weak hydrogen ingestion at $\lhe = 4.28$ during the second thermal pulse.
As a result, we obtain $\abra{C}{H} = -3.94$, $\abra{N}{H} = -1.87$,
and $\abra{O}{H} = -4.62$ after the He-FDDM-A.
However, the entropy barrier at the core is still too small to prevent a hydrogen ingestion.
For this model, therefore, we find a second He-FDDM-A in the fifth pulse where $\lhe = 7.15$, which eventually raises the surface CNO abundances become rich enough to prohibit further events of hydrogen ingestion.

\begin{figure}
\includegraphics[width=84mm]{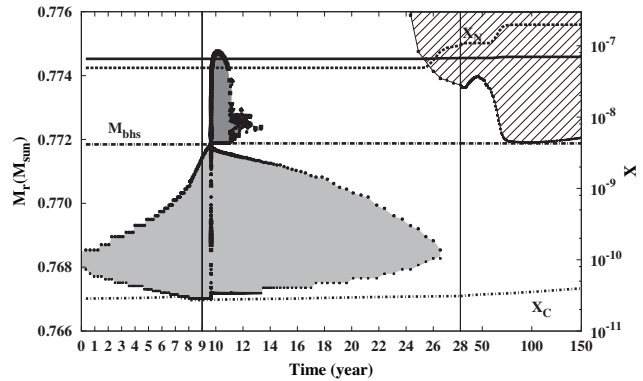}
\caption{
Same as in Fig.~\ref{fig:hefddmaa}, but for the sixth pulse of the model of $M = 3 \msun$ and $Z = 0$.
Surface abundances are shown in units of mass fraction.
The location of $M_{1}$ is omitted in this figure since it is almost identical to those of $M_{{\rm bhs}}$.
}
\label{fig:hefddmab}
\end{figure}

For $M = 3 \msun$ with $Z = 0$, we find a different mechanism of dredge-up during the sixth thermal pulse as well as the hydrogen ingestion.
In this model, the convective instabilities take place at the hydrogen-burning shell during the sixth pulse after several events of weak hydrogen ingestion.
Figure~\ref{fig:hefddmab} shows the appearance of convective zone at the bottom of the hydrogen burning shell and dredge-up in the sixth pulse.
The convective zone appears in the shell where the bottom of convection reaches down to the shell of $X \sim 10^{-5}$, which is close to the bottom of hydrogen-containing shell as shown in Fig.~\ref{fig:hefddmab}.
The reason for the appearance of convective instabilities is discussed below.
This hydrogen-burning driven convection grows in mass as much as $\pow{2.29}{-3} \msun$, and finally contains hydrogen, with $X = 0.59$.
For this case, nuclear products in the hydrogen burning shell are brought up to the surface as the convective envelope deepens.
Since the dredge-up does not penetrate into the helium-burning layer, the resultant surface abundances of CNO elements are not so large compared with the case of the He-FDDM.
We find $\xc = \pow{7.0}{-8}$, $\xn = \pow{2.0}{-7}$, and $\xo = \pow{4.6}{-9}$ after the dredge-up in the sixth pulse.
This result is different from the models of $3 \msun$ with $Z = 0$ computed by \citet{Siess2002} and \citet{Campbell2008b}.

We may conclude that the dredge-up by the hydrogen convective instability has nothing to do with the hydrogen mixing into the helium flash convection.
The computations of this model is terminated at the seventh pulse due to technical difficulties in the numerical computations.
It is to be noted, however, that the hydrogen ingestion events are still in a growing phase along with the number of pulses.
We expect therefore that the further computations of the model may increase the surface abundances because CNO abundances do not exceed the mixing limit of [CNO/H] $\sim -2.5$, and we may classify this model as He-FDDM-A.
\citet{Campbell2008b} find the mixing events and dredge-up for the same mass and metallicity.

We have encountered two types of convective instability in the hydrogen burning shell.
The first type is found for the 2 or $3 \msun$ models with $\feoh \la -5$ including $3 \msun$ with $Z = 0$ as mentioned above.
Similar results are reported for the models of $4 \mathchar`- 6 \msun$ with $Z=0$ by \citet{Chieffi2001} and for the models of 2 and $3 \msun$ with $Z=0$ by \citet{Siess2002}.
In these models, the temperature of the hydrogen-burning shell is high enough to produce CN-cycle catalysts during the stable helium shell burning on AGB.
Then, convection appears at the very bottom of the hydrogen-burning shell only with $X \la 10^{-5}$ initially.
The convection is driven by the active CN cycle for which the temperature dependence of energy generation rates is large compared with \pp\ chains.
It grows and engulfs more hydrogen into the convection and finally forms an abundance discontinuity between the hydrogen-burning shell and the helium core.
The formation of this discontinuity occurs during the decay phase of the helium flash where $L_{\he} \sim 10^4 \lsun$.
However, we could not find any contact of helium-flash convection with this hydrogen.
After the end of the helium flash, the hydrogen discontinuity disappears quickly due to the hydrogen shell burning at its bottom.
Due to the small expansion of the hydrogen-burning layer, the convective envelope slightly dredges up materials to the surface, but the surface abundances are almost unchanged.
The hydrogen-burning luminosity by this convective burning is negligible compared with the helium-burning luminosity.
Note that this type of convective instability is driven by the same mechanism as the He-H flash in low-mass Pop.~III stars caused by the strong temperature dependence of triple-$\alpha$ reaction rates \citep{Fujimoto1990,Suda2007}.

The second type of convective instability is found in $M \ga 4 \msun$ models.
This type of convection is caused by the effect of opacity just after the disappearance of helium-flash convection.
Since the convective shell appears in the middle of the hydrogen-burning shell, the profile of the hydrogen-burning shell is smooth at its bottom.
Although we find the convective instability in every thermal pulse, it does not cause any effects on the evolution and the surface abundances.
In the models of this mass range, convective instabilities do not occur at the bottom of hydrogen-burning shell because the CNO abundances are decreased by the penetration of surface convective zone at the second dredge-up.

\subsection{No hydrogen ingestion events}

For Case III models in Tab.~\ref{tab:case}, which are designated by open triangles in Fig.~\ref{fig:model}, the helium flash convection never touches the hydrogen-containing layer.
The final fate of Case III evolution is not well established and still controversial because of the uncertainty in the prescription of the mass loss rate.
\citet{Gil-Pons2007} discusses the final fate of these models by comparing the time for the core to approach Chandrasekhar mass and the time to expel the whole envelope using their adopted prescription for mass loss.
They conclude that the former is much smaller than the latter and that the stars in this mass range are likely to end their lives as carbon detonation/deflagration supernovae.
The models that ignite the carbon burning in the degenerate core finish their lives by exploding the whole star.
\citet{Tsujimoto2006} insist that there are three metal-poor stars showing the evidence of the thermonuclear supernovae of $4-7 \msun$ stars with very low metallicity ($\feoh < -4$).
They argue that these stars are formed from gas polluted by the remnant of the carbon deflagration supernovae evolved from the AGB stars whose cores approach Chandrasekhar limit due to their inefficient mass loss.

For more massive stars, carbon burning develops after the second dredge-up.
The border of the occurrence of carbon burning is shown in Fig.~\ref{fig:model}.
We terminated our computations when the carbon burning drives convection to appear.
This border is defined by $\mup$ as the critical initial mass above which carbon burning starts in the centre of the CO core without explosion.
The value of $\mup$ becomes $8 \msun$ for $\feoh \geq -4$, $9 \msun$ for $\feoh = -5$, and $8 \msun$ for $Z = 0$ in our models.
The models in this region can become white dwarfs for primordial stars as they lose their envelope through mass loss after evolving to the so-called ``Super AGB'' \citep{GarciaBerro1994,Ritossa1996}.
These models have O, Ne, and Mg cores and undergo the thermally pulsing phase of the hydrogen and helium burning \citep{GilPons2005}.
For $Z = 0$ and $5 \leq M / \msun \leq 10$, \citet{Gil-Pons2007} followed
the carbon burning and super AGB phase with and without convective overshooting.
They find no hydrogen engulfment by the helium shell flashes, which is consistent with ours and also with FII00.
They also find $\mup$ of $7.8 \msun$ without overshooting.

\section{Discussion}
In this section, we discuss the possible application of our computational results to the observed characteristics of EMP stars, starting with the discussion about model uncertainties and the comparisons with previous works.

\subsection{Model uncertainties}
We examine the uncertainties of our numerical models by changing the values of mixing length parameter for $6 \msun$ with $\feoh = -3$.
We compute several models for different $\amlt = 1.0 - 3.0$ to check whether the mixing length parameter affects the efficiency of the TDU event because the model is close to the boundary of the occurrence of the TDU events as seen in Fig.~\ref{fig:model}.
The model computed with $\amlt = 1.5$ never produced TDU events up to the 55th thermal pulse.
We re-compute the models of $54 \mathchar`- 60$th thermal pulse (over the duration of $> 10^{4}$ years which is enough time to readjust the envelope) by changing the value of $\amlt$, but do not encounter the deep dredge-up to enrich carbon in the envelope.
The change of mixing length parameters by a factor of a few seems not to change the efficiency of dredge-ups.
If we take into account significant amount of overshooting, the threshold for dredge-ups may be lowered.
\citet{Lau2009} already discussed this and obtained differences from their fiducial models within expectation.
Therefore, we will not discuss here about overshooting since unfortunately, we do not have any plausible methods for calibrating the values of overshooting as well as the mixing length parameters for AGB models at low-metallicity.

We also computed a $2.0 \msun$ model with $\feoh = -4$ with $\amlt = 3.0$ from the zero-age main sequence to see the total quantitative difference from our fiducial model of $\amlt = 1.5$.
This model has almost the same mass of helium core at the beginning of the thermal pulse as our fiducial model.
We encounter hydrogen ingestion at the 3rd, 4th, and 5th thermal pulses as in the fiducial model.
We obtain slightly different helium core mass at the 5th thermal pulse where we find He-FDDM-A, and hence, a small difference in the surface abundances after dredge-up within a few factor.
We may conclude that the dependence of mixing length parameter is weak.

Before closing this subsection, we comment on the hot bottom burning (HBB).
In our models, we do not find any signature of HBB episodes.
The temperature at the bottom of convective envelope always remains lower than $\pow{3}{7} K$ irrespective of stellar mass.
\citet{Campbell2008b} and \citet{Siess2002} insist that they find the HBB for the models of $\geq 2$ or $3 \msun$ of zero metallicity.
On the other hand, \citet{Lau2009} do not find HBB for $M \leq 3 \msun$, while they find hot third dredge-up for the models of more massive star, typically $M \geq 4 \msun$.
The hot third dredge-up is the efficient dredge-up of the former helium convective shells after the thermal pulse by the convective envelope where CN cycles operate at its bottom.
The reason for the efficient HBB at lower initial metallicity should be related to the larger temperature of the hydrogen-burning shell than in the more metal-rich counterparts because of the lack of pristine CNO elements.
However, the situation can be changed if He-FDDM occurs in the early phase of thermal pulses and enriches the CNO elements in the surface as much as those of metal-rich populations.

The surface abundances of EMP models at the AGB play an important role when we consider the origins of the most iron-poor stars currently known.
The efficient HBB events are not supported by the observations of EMP stars in the view of the dearth of nitrogen enhanced EMP stars relative to carbon-enhanced EMP (CEMP) stars.
In the current observations, only 10 \% of CEMP stars show $[\textrm{N}/\textrm{C}] \ga 0.5$ (T. Suda et al. 2010, in prep.).
The reason for the absence of HBB events is still an open question \citep{Masseron2009}.
In order to understand the difference in the operation of HBB, we have compared models with different stellar evolution codes (Suda et al 2010, in preparation).
As far as we check the details of models, a difference is likely to be due to the difference in numerical schemes.
The detailed discussion about this topic is beyond the scope of this paper because we mainly focus on the mixing events and dredge-up.
However, it should be noted that the final surface abundances presented here can be affected by HBB if it occurs during the TPAGB phase.

\subsection{Comparisons with previous works}
As mentioned in \S 1, several previous works reported the abundances after the He-FDDM events, which can be compared with our results shown in Table~\ref{tab:hefddm}.
Table~\ref{tab:compare} summarize the previous results of computations of low- and intermediate-mass stars at low metallicity except for LST09 who computed the models in the mass range of $1 \leq M / \msun \leq 7$ and the metallicity range of $-6.3 \leq \feoh \leq -2.3$ but give the resultant abundances only in their figures.

\begin{figure}
\includegraphics[width=84mm]{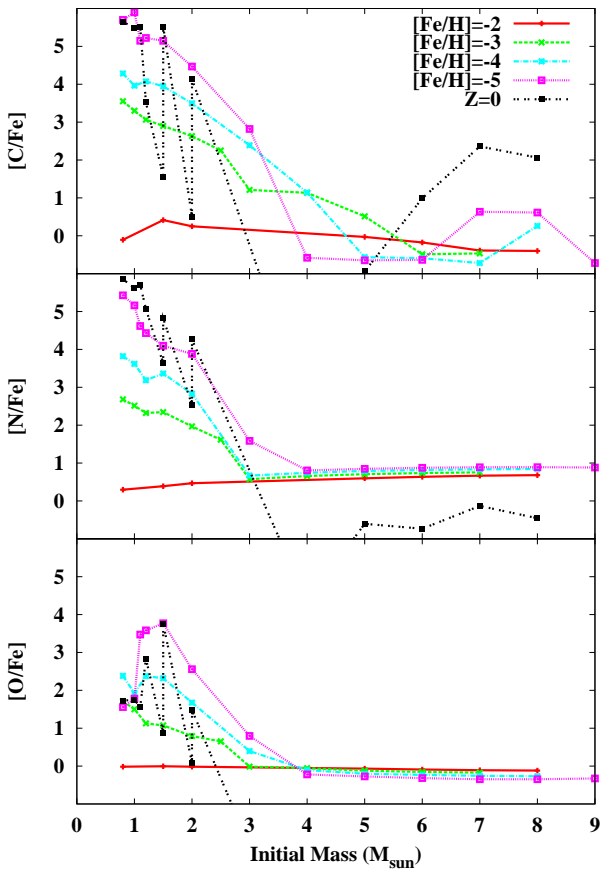}
\caption{
Final surface CNO abundances of our models at the AGB phase as a function of initial mass.
Our computations are terminated after several thermal pulses.
For Cases I, II, and Case \iip\ models, the final surface abundances are dominated by the He-FDDM events.
For $Z = 0$ models, we assume that the model surfaces are contaminated additionally by metals with $\feoh = -5.5$ to compare with the observed HMP stars.
For $1.5$ and $2 \msun$ models with $Z = 0$, we plot the results for 1st and 2nd event of the He-FDDM-A.
}
\label{fig:abn}
\end{figure}

\input{Table4.tex}

For low mass stars, most of the previous works find hydrogen ingestion into the helium-flash convection and revealed the huge enrichment of CNO abundance in the surface.
These abundances are consistent within a factor of 2 or 3 except for the oxygen abundance.
The large oxygen abundances by \citet{Picardi2004} and by \citet{Campbell2008b} (hereafter CL08), compared with ours and probably with those of \citet{Schlattl2002} is due to the neutron recycling reactions, $\nucm{12}{C} (n, \gamma) \nucm{13}{C} (\alpha, n) \nucm{16}{O}$, as discussed in \citet{Nishimura2009}.  
In our computations, we did not follow the neutron capture reactions, and hence, oxygen is produced solely by $\nucm{12}{C} (\alpha, \gamma) \nucm{16}{O}$.
Accordingly, for the He-FDDM-R and the He-FDDM-A in low-mass stars, more than an order of magnitude differences arise because of an inefficient oxygen production by $\nucm{12}{C} (\alpha, \gamma)$ at small carbon abundance in the helium flash convective zone.
However, the differences narrow for the He-FDDM-A since the oxygen production becomes efficient rapidly with increasing carbon abundance \citep[see e.g.,][]{Suda2004}.

With regard to the parameter ranges of the He-FDDM events, CL08 find He-FDDM-R (which they call Dual Core Flash) in the $0.85 \msun$ model but not in $1 \msun$ model for the metallicity of $\feoh < -5.45$.
This threshold metallicity for $1 \msun$ is lower than our result.
However, the general trend is the same, i.e., the proton ingestion is likely to take place for models that are less massive and more metal-poor.
This can be seen from the dotted lines in Fig~\ref{fig:model} that show the largest hydrogen abundance in the hydrogen-containing layer that the helium-flash convection has reached.
As for He-FDDM-A, the upper metallicity boundary of the occurrence ($\feoh \simeq -2.5$) is somewhat larger than obtained by CL08 and LST09.
CL08 find He-FDDM-A events (which they call Dual Shell Flash) for the metallicity, $\feoh < -3$ and $< -4$ in their models of $M < 2 \msun$ and $M \geq 2 \msun$, respectively.
The boundary metallicity of hydrogen flash in LST09 is almost the same as in CL08.
As for the mass boundary of He-FDDM-A, our result presented in Fig~\ref{fig:model} is similar to, though slightly larger than, the results both by CL08 ($M \leq 2 \msun$ for $\feoh = -4$ and $M \leq 3 \msun$ for $\feoh \leq -5$) and by LST09 ($M \leq 2 \msun$ for $\feoh \leq -4$), as well as to that of FII00.
On the other hand, our mass and metallicity threshoulds for He-FDDM-A agree well with those of \citet{Iwamoto2004}, in contrast to the discrepancy with those of CL08 and LST09.

Fig.~1 of CL08 shows the time evolution of convective zones during the He-FDDM-R event for $1 \msun$ with $\feoh = -6.5$. 
Our comparative model of $1 \msun$ with $Z = 0$ gives quantitatively similar results.
In our model, core helium flash ignites at slightly smaller core mass ($M_{1} = 0.499 \msun$ vs. $M_{1} = 0.51 \msun$ in CL08) in slightly outer mass shell ($M_{r} = 0.317 \msun$ vs. $M_{r} = 0.295 \msun$ in CL08).
Accordingly, the pressure in the helium burning shell is smaller, and hence, the helium flash is weaker in ours than in CL08.
The mass in the hydrogen-flash convective zone is smaller in our model than in CL08, ($0.09 \msun$ vs. $0.1 \msun$), though the difference reduces because of larger radius at the shell of ignition.
The duration of the hydrogen-flash convection is shorter ($\simeq 500$ versus $\simeq 1000$ years).
On the other hand, the dredge-up mass is larger in our models than in CL08 by a factor of $\sim 2$, since the helium flash, and hence, the hydrogen flash, occur in the shells closer to the hydrogen-burning shell.
Such small differences may readily arises from the difference in the numerical treatment such as the treatment of spacing and rezoning of mesh points.
Slight differences in temperature can cause considerable changes in the condition of igniting helium flashes because of the strong temperature dependence of helium-burning rate at the ignition.

For LST09, we may compare the models of our $1.5 \msun$ at $\feoh = -3$, $3 \msun$ at $\feoh = -5$, and $4 \msun$ at $Z = 0$ with their models of same masses at $\feoh = -3.3$, $-5.3$, and $-6.3$.
For all of these models, they have larger helium core mass at the beginning of thermal pulses on AGB phase.
The largest difference is $0.07 \msun$ ($13\%$ of the core mass) between the
$1.5 \msun$ models, while the difference grows smaller to be only $0.01$ - $0.02 \msun$ for more massive models like 3 and $4 \msun$.
The larger core mass may be one of the reasons that they find a boundary for He-FDDM at smaller initial mass and metallicity.

For $1.5 \msun$ models, our models have larger mass in the helium-flash convection during helium shell flashes than those of LST09 ($0.035 \msun$ compared with $0.024 \msun$) due to the smaller masses of carbon-oxygen core.
Both meet with the He-FDDM-A event at the third pulse on the TPAGB.
In spite of the difference in helium core mass, the maximum helium-burning luminosity is almost the same in both models ($\log (L_{\textrm{He}}^{\textrm{p}} / \lsun) = 7.6$, see Tab.~\ref{tab:hefddm}).
The dredge-up mass by this event is two times larger in our model ($0.018 \msun$) than in LST09 ($0.009 \msun$), while the dredge-up efficiency is smaller in ours (80 \% of the former hydrogen-flash convective zone) than in theirs (almost 100 \%).
The maximum hydrogen-burning luminosity is much larger in our model.
Our model reaches $\log (L_{\textrm{H}}^{\textrm{p}} / \lsun) \sim 10$, while they have only $\sim 7$.
For $3 \msun$ models, the growth of thermal pulses is similar in both models in their height and interval.
However, they find a so-called ``carbon ingestion'', as discussed later, while we do not.
For $4 \msun$ models, the result is quite different in the sense that they find deep convective envelope well below the bottom of the hydrogen-burning shell (they call hot third dredge-up) after the seventh thermal pulse.
The bottom of convective envelope comes close to the carbon oxygen core in their model.
In contrast, we only find ordinary third dredge-up at tenth thermal pulse for this model.
This is the same phenomenon as ``hot dredge-up'' found in $5 \msun$ models of \citet{Herwig2003}.
We still do not know why such quite different efficiency of the TDU is obtained by different stellar evolution code.
From the published data, it is difficult to pinpoint the critical reason for differences between models because of a lack of comparable data.

For intermediate mass stars with M $\geq$ 4 \msun, \citet{Chieffi2001} also demonstrate that
``carbon ingestion'' occurs for AGB models of $Z = 0$ and 4 - $6 \msun$.
This is driven by the hydrogen-burning instability during the thermal pulse, which causes the penetration of the hydrogen convective zone into the underlying carbon-rich layer, assisted by the induced overshooting.
An injection of hydrogen into the carbon-rich layer gives rise to a hydrogen flash in carbon-rich layer to deepen the convective envelope in mass, leading to the enhancement of CNO elements in the surface convection as much as $\xcno \sim \pow{4}{-6}$.
It is suggested that additional mixing, such as increased overshooting, leads to the enrichment of surface CNO elements to values like those in Pop.II AGB models.
\citet{Siess2002} found and named ``carbon injection'' similar to that found by \citet{Chieffi2001} and detected the inward extension of convection by the hydrogen shell burning during thermal pulses for $Z = 0$ and $M = 1, 1.5, 2, 3, 4$, and $5 \msun$ models.
\citet{Herwig2003} also finds the hydrogen convective episode at the bottom of the hydrogen-burning shell for $5 \msun$ model with $Z = 0$, although it does not cause the carbon ingestion.

The different results for the carbon ingestion are ascribed to the different treatment of convective mixing.  
The carbon ingestion events are found in stellar evolution codes that mix the layers over the convective boundary with the discontinuity of chemical composition.
Since the treatment of convective overshooting with such large difference in the molecular weight is not well established, and since the evolution and nucleosynthesis in these metal-poor models are sensitive to the treatment of mixing and burning in the code, numerical details should be investigated more carefully.

\subsection{Comparisons with observed EMP stars}
If we compare the abundance of models with those of observed stars directly, we should adopt the models with mass $M \la 0.8 \msun$ because most of the known EMP stars have long lifetimes equal to or longer than the age of the universe.
On the other hand, we can compare the abundances of models with $M > 0.8 \msun$, if we assume that binary mass transfer or wind accretion occurs in a binary system consisting of a massive primary and a low-mass secondary whose mass is $\sim 0.8 \msun$.
We have emphasized the role of binaries as addressed in the previous works \citep{Suda2004,Komiya2007,Komiya2009a}.
In this case, however, it seems difficult to constrain how much mass is accreted onto the surface of the secondary in the EMP binaries.
This is true even if the binary parameters are known from observations since the evolution of stars in the close binary systems critically depends on the angular momentum loss from the systems, which is yet to be well constrained both theoretically and observationally.
Accordingly, in this work, we do not consider the details of the binary evolution.

Figure~\ref{fig:abn} summarizes the surface CNO abundances of our low- to intermediate-mass models.
Here we assume that $Z = 0$ models are polluted with metals by interstellar accretion to as metal-rich as $\feoh = -5.5$, roughly comparable with the abundances of hyper/ultra metal-poor (HMP/UMP) stars, defined as having metallicity below $\feoh < -4.5$.
For $M = 1.5$ and $2 \msun$ models with $Z = 0$, we add the results of the two He-FDDM-A events.
As can be seen in the figure, we obtained various trends for the ratio of CNO elements, which are useful indicators for the origins of observed EMP stars.
As for oxygen, our results set a lower bound for the enrichment since our models do not take into account the oxygen production due to neutron recycling reactions. 
Nevertheless, our results suggest that the oxygen production by $\nucm{12}{C} (\alpha, \gamma) \nucm{16}{O}$ alone can produce a fairly large oxygen enrichment to explain some of the observations.
The oxygen enrichment in EMP stars is discussed by \citet{Nishimura2009} in detail with the oxygen production taken into account in the helium-burning shell both through the $\alpha$ capture of \nucm{12}{C} and the neutron recycling reactions.

\begin{figure}
\includegraphics[width=84mm]{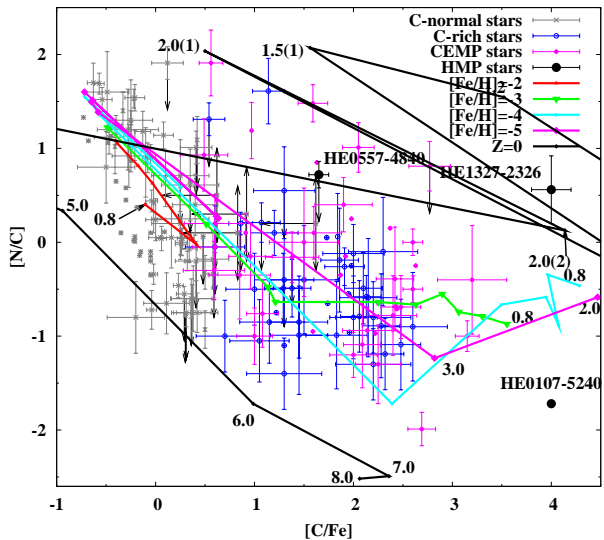}
\caption{
Comparison of the carbon and nitrogen enhancement obtained by the model computations with the observed abundances for EMP and HMP/UMP stars. 
The present results are shown by the same symbols connected by the same types of lines as in Fig.~\ref{fig:abn}.
Numerals attached to some of points indicate the model masses.
The labels with ``(1)'' and ``(2)'' denote the results after the first and second event of hydrogen ingestion, respectively.
The observed abundances are taken from the SAGA database \citep{Suda2008} and divided into three groups; C-normal stars for $\cfe < 0.5$, C-rich stars for $\cfe \geq 0.5$ and $\feoh > -2.5$, and CEMP stars for $\cfe \geq 0.5$ and $-4.5 \leq \feoh \leq -2.5$, and HMP/UMP stars for $\feoh < -4.5$.
}
\label{fig:cno}
\end{figure}

Figure~\ref{fig:cno} shows the observed C and N abundances, taken from the SAGA database \citep{Suda2008}.
For models after He-FDDM events, we obtain the range, $-2 \la \cnratio \la 1$.
If these models experience the subsequent TDU events, only carbon can be enhanced, and hence, the data go toward bottom right in Fig.~\ref{fig:cno}.
In general, we find extremely large enhancement of carbon and nitrogen by He-FDDM-R, because of relatively large mass in the flash convective zone and of relatively small envelope mass.
For He-FDDM-A, we find small variations in abundance ratios for between $\feoh = -3$ and $-5$, while $Z = 0$ models have different ratios depending on initial mass.
It should be noted that our model computations are terminated after several thermal pulses without considering mass loss.
The final surface abundances depend on the mass loss history of progenitor stars, although the surface abundances are strongly influenced by the He-FDDM events.

As for the three HMP/UMP stars known to date, HE0107-5240, HE1327-2326, and HE0557-4840, we may suggest that their progenitors are the low-mass members of binary systems whose surfaces have been polluted with the envelope matter ejected by the primary stars evolving to AGB.  
Regardless of the details of binary mass accretion, the CNO abundances of HE0107-5240 \citep{Christlieb2002,Christlieb2004b} may be consistent with those of $1.5 - 3 \msun$ models with $\feoh = -5$ or $Z = 0$, if we take into account the carbon enrichment by the TDU subsequent to He-FDDM, as discussed in \citet{Suda2004} and \citet{Nishimura2009}.
For HE1327-2326, $M = 1.5 - 2.0 \msun$ with $Z = 0$ model agrees well for carbon and nitrogen abundances derived by \citet{Aoki2006b} without recourse to TDU.
The oxygen abundance derived by \citet{Frebel2006b} is explicable in terms of neutron recycling reactions in the helium-flash convective zones as discussed in \citet{Nishimura2009}.
The abundance pattern of HE0557-4840 may be the result of the TDU event as discussed in \citet{Nishimura2009}.
The model of $M = 3 \msun$ with $\feoh = -5$ seems to agree with the abundance trend of CNO elements, but we cannot exclude the possibility of He-FDDM-A with weak hydrogen ingestion at $M = 3 \msun$ model with $Z = 0$.
For this star, we have to wait for the abundances for nitrogen and oxygen since \citet{Norris2007} give only the upper limits.
These interpretations are shown to be compatible with the observed abundances for other elements such as Na, Mg, and Al \citep{Nishimura2009}, and will be discussed for s-process elements in the subsequent paper (Yamada et al.~ in preparation).
Of course, the detailed comparisons require the more sophisticated modelling of binary evolution considering the mass loss and mass accretion history and the element mixing between the accreted and envelope matter in the secondary of the binary system.  
Finally, we note that for all of these stars, the self-pollution by the He-FDDM-R in $\sim 0.8 \msun$ stars should be rejected because of too large enhancement of carbon and nitrogen such as $\cfe \sim \nfe > 5$ \citep[see also,][]{Picardi2004,Campbell2008b}.

\subsection{Other modifications of surface abundances}
So far, we have focused on the hydrogen ingestion and He-FDDM, and in this section, we discuss the other events that may affect the surface abundances.
The main driver of changing surface abundances is the surface convection in the envelope that deepens in mass to dredge up the nuclear products of the hydrogen and helium-burning.
There are three ocasions that with regards to the notation of the dredge-up mechanisms,
we adopt the following convention that is commonly used; the 1st dredge-up occurs at the beginning of the red giant branch, the second dredge-up occurs at the beginning of the TPAGB phase, and the 3rd dredge-up occurs during double shell burning.

\begin{figure}
\includegraphics[width=84mm]{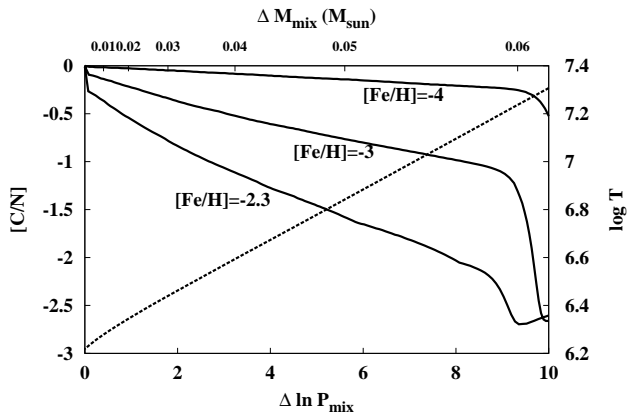}
\caption{
Abundance profile of carbon to nitrogen ratio in the envelope of $0.8 \msun$ models for various metallicities as a function of the logarithmic pressure ratio, $\Delta \ln P_{\rm mix} = \ln (P/P_{\rm bc})$, between the pressure, $P$, at the shell and the $P_{\rm bc}$ at the bottom of surface convection (solid lines).
Models are taken at the maximum depth of surface convection during the first ascent on the giant branch.
Dotted line denotes the temperature for the model of $\feoh = -3$.
The axis of abscissa in the top gives the mass contained between the shell and the bottom of surface convection for the model of $\feoh = -3$.
}
\label{fig:mixed}
\end{figure}

For the EMP stars, the 1st dredge-up in $0.8 \msun$ stars only slightly changes the surface abundances because of the shallower surface convection when compared with the more metal-rich models.
Our models show the change of helium abundance by less than two percents, and also, changes in the CNO abundances by less than one percent, regardless of metallicity.
On the other hand, \citet{Spite2005} find some C-poor and N-rich stars among EMP giants and named them ``mixed'' stars. 
The mixed stars are reported to have typically [C/N] $\sim -1.5$, while ``unmixed'' stars have [C/N] $\ga 0$.
This can be brought about by the internal mixing during the first ascent on the red giant branch, while also possible is pollution by AGB stars in the binaries, as discussed by \citep[see also,][]{Spite2006}.
In the former case, the conversion of carbon into nitrogen in the envelope requires extended mixing (so-called ``extra mixing'') in the radiative zone below the surface convection.
Figure~\ref{fig:mixed} shows the profile of abundance ratio $\abra{C}{N}$ in the radiative zone as a function of the depth of the shell from the bottom of surface convection, measured in pressure, $\Delta \ln \pmix \equiv \ln (P / P_{\rm bc})$, where $P$ and $P_{\rm bc}$ are the pressures at the shell and at the bottom of surface convection, respectively, for various metallicity.
The observed ratios of $\abra{C}{N} (\simeq -1.5)$ are realized in the shells deeper by 5 pressure scale-heights for the model of $\feoh =-2.3$ to by more than10 pressure scale-heights for the model of $\feoh = 4$, depending on the metallicity; 
the depth corresponds to $\sim 0.06 \msun$ in mass for the model of $\feoh =-3$.
This may give an estimate of the depth that extra-mixing has to reach in order to realize the abundance changes observed for ``mixed'' stars during the RGB evolution, though the actual degrees of abundance changes also depend on the time scale of extra mixing.

The characteristics of the second dredge-up are summarized in Table~\ref{tab:evo}.
The effect of the second dredge-up is significant in the helium enhancement for lower metallicity.
For $Z = 0$, the helium mass fraction exceeds $0.3$ for the initial mass of $2.5 \msun$, while this is the case for $M \geq 6 \msun$ and $\feoh \geq -4$.
The larger enrichment of helium for lower metallicity comes from a less steep hydrogen profile in the envelope.
Because of the lack of initial CNO elements, \pp\ chains are dominant in the outer part of the hydrogen burning shell, which forms a gradual decrease of hydrogen abundance from the surface to the middle of the hydrogen burning shell due to the smaller temperature dependence of nuclear reaction rates in \pp\ chains than in CNO cycles.
As a result, helium is abundant in the outer hydrogen-burning shell and is easily enhanced at the surface by the deepening of the convective envelope.

Our models show the dependence of the occurrence of TDU on initial mass
and metallicity.
For the low-mass stars, the evolution after the He-FDDM events is similar to that of more metal-rich models because the increase in the abundance of CNO elements in the envelope activates the CNO cycle reactions and raises the hydrogen burning rate.
In most of our models, we stopped computations after the deep mixing and do not find the TDU because of the decreased helium core by the dredge-up.
In an effort to find the TDU, however, we follow the evolution after the He-FDDM for $2.0 \msun$ with $\feoh = -4$ and $1.5 \msun$ with $\feoh = -2$ and $-4$.
We find the TDU after the 35th pulse for $2 \msun$ when the mass of helium core grows to be $0.780 \msun$.
On the other hand, we failed to find the TDU for $1.5 \msun$ model with $\feoh = -4$ after the He-FDDM, although we follow the computation until the 49th thermal pulse with helium core mass of $0.789 \msun$.
For the model of $1.5 \msun$ with $\feoh = -2$, we find the TDU at 29th pulse when $M_{1} = 0.742 \msun$.
This seems consistent with the result that the TDU occurs for $M \geq 1.5 \msun$ by \citet{Lattanzio1987} ,although their models have $\feoh > -1$.

\begin{figure}
\includegraphics[width=84mm]{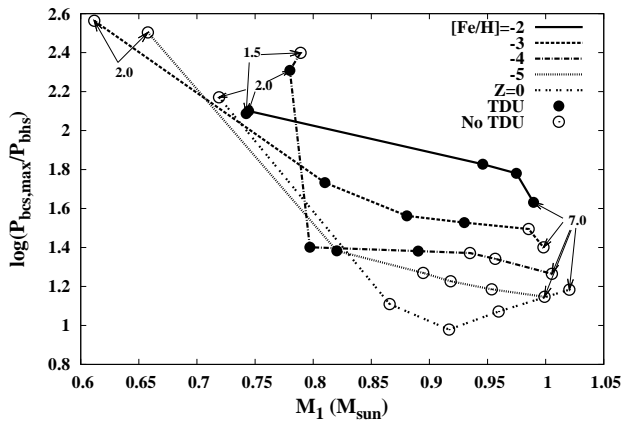}
\caption{
Thickness of the helium-flash convective zones in pressure as a function of core mass if a third dredge-up episode takes place after the thermal pulse (filled circles).
If the models never encounter the TDU, we take the final models of the TPAGB (open circles).
The thickness of the helium-flash convective zones is measured by the ratio of maximum pressure at the bottom of convective zone during the thermal pulse $\ln P_{\rm bcs,max}$ to the bottom of hydrogen-containing shell at that time $P_{\rm bhs}$.
Lines connect the results for the models with the same metallicity.
For some models, the initial masses are labelled next to the point.
}
\label{fig:tdu}
\end{figure}

For more massive stars, the efficiency of TDU turns to decrease, and hence, the change of surface abundances becomes smaller, as seen in Fig.~\ref{fig:model},
As for the metallicity dependence, the efficiency of TDU decreases with decreasing metallicity.
These trends can be interpreted as the strength of the helium shell flashes because strong flashes expand the envelope, which enhances the efficiency of the TDU.
In Figure~\ref{fig:tdu}, we show the strength of the helium shell flashes for the computed models just before the thermal pulses prior to the first TDU event or the final thermal pulses of our computations without any TDU events.
The figure shows the ratio of the maximum pressure at the bottom of the helium-flash convective zone to the pressure at the bottom of the hydrogen-containing shell as a function of helium core mass.
First, there exists a critical core mass to encounter the TDU events, which is around $0.74 \msun$ and may depend on the pressure ratio and metallicity.
Another critical value to drive the TDU is the pressure ratio shown in the figure, which slightly increases with increasing helium core mass.
The strength of the shell flashes measured by the pressure is first discussed in \citet{Sugimoto1978}, who define the proper pressure that takes into account the flatness of the helium-burning layer before the ignition.
Using the proper pressure of the thermal pulses, we obtain the same correlation with the occurrence of the TDU events as in Fig.~\ref{fig:tdu}.

One should note that the efficiency or the occurrence of the TDU events depends on the status of the envelope as well as that of the helium shell flashes.
One of the important factors that determine the physical status of the envelope is the entropy of the hydrogen-burning shell.
It affects the compression of helium layer in the core for the models with $Z > 0$ through small hydrogen burning rate.
For $Z = 0$, the temperature in the hydrogen-burning shell directly determines the thermal state of the helium core as discussed in \citet{Suda2007}.
Therefore, the condition for the occurrence of the TDU events at $Z = 0$ can be different from other models with finite metallicity.
Another possible factor to determine whether the TDU occurs or not is the inclusion of overshooting at the bottom of helium-flash convective zones as discussed by \citet{Herwig2000}.
If it were introduced in our models of extremely metal-poor and metal-free stars, the strength of the helium flashes may increase enough to drive the TDU events.
The detailed discussion about the occurrence of the TDU events can be even more complicated than the discussion here, and furthermore, we have not yet the precise prescription for the strength of overshooting, which needs elaboration in future works.

It is to be noted that the carbon abundance is enhanced for $M = 6$ - $8 \msun$ with $Z = 0$ models as seen in Tab.~\ref{tab:evo} and Fig.~\ref{fig:abn}.
This enhancement is not due to the TDU but the second dredge-up.
In $Z = 0$ models, helium-rich layer has a high temperature, i.e., $\log T \approx 7.8$ so that 3-$\alpha$ reactions take place during the second ascent on the red giant branch.
These carbon-rich shells are dredged up to the surface just before the thermal pulse phase.
As a result, the surface abundance of these models become carbon-rich compared with other light elements, although the absolute abundances are as low as [C/H] $\sim -3$ (see also Table~\ref{tab:evo}).
\citet{Lau2009} insist that the 6 and $7 \msun$ models with $Z = 10^{-8}$ undergo TDU events, while $5 \msun$ model with the same metallicity does not.
They also insist that this irregular dependence of the third dredge-up efficiency is due to the different CNO abundances in the hydrogen-burning shell \citep{Lau2008}.
In our models, we certainly find carbon enhancement by the second dredge-up in $M \geq 7 \msun$ models for $\feoh \leq -4$ and $M \geq 5 \msun$ models for $Z = 0$.
However, we still do not find any TDU events in these models.
As seen in Fig.~\ref{fig:tdu}, the strength of shell flashes for these models are still below the critical values to drive the TDU event.
The effect of carbon enhancement in the hydrogen-burning shell can be seen only for $Z = 0$ as the increase of pressure ratio with increasing initial mass for $M \geq 5 \msun$ in Fig.~\ref{fig:tdu}.

\subsection{Horizontal branch EMP models}
\begin{figure}
\includegraphics[width=84mm]{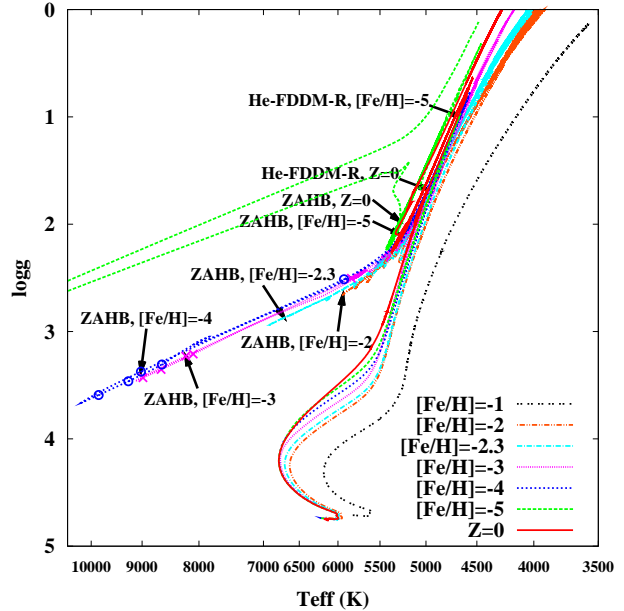}
\caption{
Stellar evolution tracks on the $\log g$ - $\teff$ diagram for $0.8 \msun$ with various metallicity.
The onset of the hydrogen-mixing event He-FDDM-R is shown by arrows for the model of $\feoh = -5$ and $Z = 0$.
The locations of zero age horizontal branch (ZAHB) are also shown for models of $\feoh \leq -2$.
Crosses and circles on the horizontal branch denote an evolutionary point of every 10 Myr starting from the ZAHB for $\feoh = -3$ and $-4$, respectively.
}
\label{fig:hrd}
\end{figure}

One of the characteristics of the EMP model is the extended horizontal branch (HB) for $0.8 \msun$ having a high effective temperature, greater than 7000K, without any modification of envelope mass and composition (See Figure~\ref{fig:hrd}).
These blue HB stars can be seen for $\feoh \la -2.5$.
The location of the horizontal branch shifts blueward as the \pp\ chains become dominant energy source of hydrogen burning.
For our models of $\feoh = -3$ and $-4$, luminosity by \pp\ chains, $L_{pp}$ is the order of $10 L_{\sun}$.
On the other hand, luminosity by CN cycles, $L_{CN}$ is 10 times smaller than $L_{pp}$ at the zero-age horizontal branch, while they become comparable for $\feoh = -2.3$.
Along with the decrease of the contribution by CN cycles, the entropy of the hydrogen-burning shell decreases.
This causes the shrinkage of the whole star through the increase of density and pressure of the hydrogen-burning shell.
The model stars of $0.8 \msun$ with $\feoh \la -2.5$ at the zero age horizontal branch have typical radii of $\la 5 R_{\sun}$.
Consequently, the effective temperature of model stars at zero age horizontal branches increases with decreasing metallicity.

Observationally, we cannot confirm the metallicity gradient along the HB in the Galactic halo because of the lack of detailed abundance analyses for HB stars of $\feoh \la -3$.
Instead, the field HB stars are used as the tools for investigating the kinematics and dynamics in the Galaxy.
Without a selection bias, blue HB stars should appear in H-R diagram and vary in \teff\ with the initial metallicity of stars if the significant modifications of surface composition, such as radiative levitation, are absent and mass loss is not significant.
Since the lifetime of blue HB stars is approximately 30 Myr (see Fig.~\ref{fig:hrd}) and is $\simeq 30$ times smaller than that of RGB stars, we expect to find at least one blue HB star per 30 RGB stars on average having comparable luminosity among stars with $-4 \la \feoh \la -2.5$.
When $\feoh \leq -5$ initially, the He-FDDM-R leads to the enhancement of carbon and nitrogen in the hydrogen-burning shell.
This does not follow the blueward extension of the horizontal branch because $L_{CN}$ becomes $\sim 100$ times larger than $L_{pp}$.
Therefore, $\feoh \sim -4$ should be the lower limit in metallicity for the absence of red horizontal branch stars.
It is known that the reduction in the envelope mass due to mass loss and the surface helium enrichment due to the extra-mixing on the RGB can promote the blue horizontal branch stars \citep[e.g., see ][and references therein]{Suda2006}. 
As suggested for the second parameter problem in the globular cluster stars, however, this may works only for the metallicity $\feoh \la -1$, 
Accordingly, the $0,8 \msun$ models with $\feoh \la -5$ can stay on the red horizontal branch since the He-FDDM-R event is likely to give rise to the surface CNO enhancement larger than $\abra{CNO}{Fe} \ga -1$ (see Tab.~\ref{tab:compare}).
The possible peculiar morphology of horizontal branch can be related to the work by \citet{Beers2007b} who suggest that the 78 percent of horizontal branch stars in the sample of HK surveys are likely to be blue horizontal branch.
This may be consistent with the present models of metal-poor stars if most of the sample stars are in the metallicity range of $-4 \la \feoh \la -2.5$, although the sample includes thick disk stars.

Finally, we comment on the blueward extension in Fig.~\ref{fig:hrd} for the model of $\feoh = -5$.
This model experiences a blue loop when $M_{1} \geq 0.67 \msun$ with thin envelope.
This is caused by the rapid growth of helium core during the horizontal branch phase after the decreased helium core ($M_{1} = 0.426 \msun$) and the large helium enhancement in the envelope ($X_{\textrm{env}} = 0.52$) by the He-FDDM-R event.
For $Z = 0$, on the other hand, core mass increases from $0.455$ to $0.599 \msun$, which does not cause a blue loop.
The lifetime of this blue loop is the order of 10 Myr and may be observable, although it depends sensitively on the depth of the dredge-up by the He-FDDM-R event.

\section{Conclusions}
We modelled the evolution of metal-poor stars below $\feoh = -2$ from the zero-age main sequence through the asymptotic giant branch phase with mass range between 0.8 and $9.0 \msun$.
The present computations focus mainly on the hydrogen mixing from the bottom of the hydrogen burning shell into the helium-flash convection at the core helium flash and/or at the helium shell flashes, depending on the initial mass and metallicity.
The condition that hydrogen mixing occurs at the core helium flash has rather complicated dependence on initial mass and metallicity, different from the previous results of \citet{Fujimoto2000} in which the computations were done with significantly different input physics from the present work.

We followed the evolution after the hydrogen-mixing event and the subsequent dredge-up during the helium shell flash phase.
For the mass range, $0.8 \leq M / \msun \leq 2$ for $-5 \leq \feoh \leq -3$ and $0.8 \leq M / \msun \leq 3$ for $Z = 0$, we find significant changes in the surface abundances after the dredge-up by the helium-flash driven deep mixing (He-FDDM).
We also find the models with the hydrogen ingestion into the helium flash convective zones without the hydrogen flash and associated deep mixing.
These models ($3 \la M / \msun \la 5$ and $-5 \la \feoh \la -2.5$) can be the origins of the so called CEMP-no{\it s} stars if {\it s}-process by the radiative \nucm{13}{C} pocket is inefficient.

Our models show the dependence of the efficiency of the third dredge-up (TDU) on initial mass and metallicity.
The efficiency of the TDU depends both on the strength of helium shell flashes and on the mass of carbon-oxygen core.
We derived the minimum thickness of the helium-burning shell required for shell flashes to drive the TDU.
The critical core mass for the TDU is $\approx 0.74 \msun$ with possible dependence on the thickness of the helium-burning shell and metallicity.
In order for the TDU to take place, models must satisfy both of these criteria.
The TDU will not take place below this critical mass even if the helium shell flash is strong enough and vice versa.
The resultant upper limit of the occurrence of the TDU is $5 \msun$ for $\feoh = -3$, $4 \msun$ for $\feoh = -4$, and $3 \msun$ for $\feoh = -5$.
We do not find any TDU events for $Z = 0$ models because of the weak helium flashes, which may be consistent with the result of \citet{Lau2009}.

This work also discusses the relevance to the observed properties of extremely metal-poor (EMP) stars.
We compare the carbon and nitrogen abundances of observed EMP stars with the surface abundances of our models.
Most of the observed CEMP stars can be explained by the modification of surface abundances during AGB phase
through the He-FDDM and the TDUs.
For the most iron-deficient stars currently known, we propose the following binary scenario for the origins of these objects by comparing the CNO abundances only:
\begin{itemize}
\item HE0107-5240: primary star is in the mass range $1.5 \la M_{1} / \msun \la 3.0$ and have undergone the He-FDDM during the thermal pulses on AGB.
This star also may have changed its surface abundances by the TDUs.
It is yet to be answered if the star is a primordial star or not.
\item HE1327-2326: primary star is in the mass range around $2 \msun$.
The He-FDDM should have taken place in TPAGB phase but not TDUs.
The model with $Z = 0$ seems to match with the observed abundances for carbon and nitrogen.
\item HE0557-4840: firm conclusion can hardly be deduced from the current observed abundances, which is less constraint.
The TDUs without the He-FDDM seems to be responsible for the abundances, while the possibility of weak hydrogen ingestion into the helium-flash convective zones cannot be excluded.
\end{itemize}

A blueward excursion of horizontal branch stars is found at the metallicity range of $-4 \la \feoh \la -2.5$.
The high effective temperatures of these models are due to the lack of CNO elements in the hydrogen-burning shell because of low metallicity when they evolve into the zero-age horizontal branch (ZAHB).
For $\feoh \la -5$, the He-FDDM occurs at the core helium flash phase, which enriches the hydrogen burning shell with CNO elements and moves the ZAHB to the red.
It is expected that one blue horizontal branch stars will be found among 30 EMP giants in the Galactic halo, estimated from the lifetime of blue horizontal branch stars in this metallicity range.
It is also expected that red horizontal branch stars in this metallicity range will be found only for $\feoh \la -5$.
In addition, we may find a metallicity gradient along the horizontal branch for Galactic halo stars or stars in metal-poor dwarf galaxies.

The detailed treatment of material mixing and nuclear burning is crucial for modelling the TPAGB phase.
In particular, for EMP models, the contact of helium-flash driven convection with the hydrogen-burning shell is very sensitive to the numerical scheme in the one-dimensional stellar evolution code.
The situation can be even more complicated if we consider the convective mixing in three-dimension \citep[see, e.g.,][]{Herwig2006}.
Since such hydrogen-mixing events severely affect the subsequent evolution, the resultant surface chemical composition will depend on the mixing length parameter and the treatment of overshooting and diffusion.
Unfortunately, there are no constraints on all of these from observations of EMP stars.
Despite these difficulties, it is worth exploring the models of EMP stars.
Further investigations of mixing in EMP models and of comparisons with observations will be provided in future works.

\section*{Acknowledgments}

We would like to acknowledge Michael E. Bennett and Raphael Hirschi for
useful comments in revising the manuscript.
We are grateful to the anonymous referee for useful comments to improve
the manuscript.
This work has been partially supported by Grant-in-Aid for
Scientific Research (18104003, 19740098), from Japan Society of the
Promotion of Science.
T. S. has been supported by a Marie Curie Incoming International Fellowship 
of the European Community FP7 Program under contract number PIIF-GA-2008-221145.

\bibliographystyle{mn}
\bibliography{mn-jour,reference}

\bsp

\label{lastpage}

\end{document}

%% file: Table1.tex
\begin{table*}
 \centering
 \begin{minipage}{140mm}
\caption{Summary of the evolutionary characteristics of stars}
\label{tab:case}
\begin{tabular}{llllll}
\hline
Case & He-FDDM$^{a}$ & TDU & Carbon$^{b}$ & Nitrogen$^{c}$ & {\it s}-process elements$^{d}$ \\
\hline
 I  & RGB & no & yes & yes & no \\
II & AGB & no & yes & yes & yes \\
\iip & AGB & yes & yes & yes & yes \\
III  & no & no & no & no & no \\
IV  & no & yes & yes & yes/no & yes$^{e}$ \\
\ivp & no & yes & yes & yes/no & yes/no$^{e,f}$ \\
\hline
\end{tabular}

\medskip
$^{a}$ Evolutionary stage where the He-FDDM occurs or not. \\
$^{b}$ Enrichment at the surface by the He-FDDM and/or the third dredge-up. \\
$^{c}$ Enrichment at the surface by the He-FDDM and/or the hot bottom burning. \\
$^{d}$ Enhancement of {\it s}-process elements by convective nucleosynthesis caused by hydrogen ingestion for $\feoh \la -2.5$ (see \citet{Suda2004},\citet{Nishimura2009} and Nishimura et al. in prep.). \\
$^{e}$ The occurrence of {\it s}-process depends on the efficiency of the \nucm{13}{C} pocket. \\
$^{f}$ The occurrence of {\it s}-process depends on the neutron production driven by hydrogen ingestion.
\end{minipage}
\end{table*}

%% file: Table2.tex
\begin{table*}
 \centering
 \begin{minipage}{140mm}
  \caption{Characteristic values before the core helium-flash or the thermally pulsating AGB phase}
  \label{tab:evo}
  \begin{tabular}{llllllllll}
\hline
Mass & $\feoh$ & $M_{1,max}^{a}$ & $M_{bc,2DU}^{a}$ & $X_{\textrm{He}}^{a}$ & $X_{\textrm{C}}^{a}$ & $X_{\textrm{N}}^{a}$ & $X_{\textrm{O}}^{a}$ \\
($\msun$) &  & ($\msun$) & ($\msun$) &  &  & \\
\hline
0.8    & -1.6       & 0.5302 & 0.5456        & 0.2449 & 7.261(-5)    & 3.730(-5)    & 6.299(-4)    \\
\hline
2.0    & -2         & 0.5689 & 0.5839        & 0.2530 & 1.617(-5)    & 3.296(-5)    & 9.745(-5)    \\
5.0    & -2         & 1.2014 & 0.9028        & 0.2928 & 1.369(-5)    & 4.587(-5)    & 8.605(-5)    \\
6.0    & -2         & 1.4382 & 0.9431        & 0.3129 & 1.309(-5)    & 5.029(-5)    & 8.181(-5)    \\
7.0    & -2         & 1.6978 & 0.9815        & 0.3300 & 1.267(-5)    & 5.379(-5)    & 7.843(-5)    \\
8.0    & -2         & 1.9709 & 1.0794        & 0.3388 & 1.269(-5)    & 5.569(-5)    & 7.663(-5)    \\
\hline
0.8    & -2.3       & 0.5291 & 0.5458        & 0.2419 & 1.275(-5)    & 1.084(-5)    & 4.984(-5)    \\
2.0    & -2.3       & 0.6232 & 0.6385        & 0.2555 & 7.814(-6)    & 1.778(-5)    & 4.768(-5)    \\
\hline
0.8    & -3         & 0.5298 & 0.5560        & 0.2404 & 2.308(-6)    & 2.590(-6)    & 9.629(-6)    \\
1.0    & -3         & 0.5356 & 0.5642        & 0.2464 & 2.234(-6)    & 2.576(-6)    & 9.745(-6)    \\
1.2    & -3         & 0.5384 & 0.5692        & 0.2518 & 2.002(-6)    & 2.743(-6)    & 9.863(-6)    \\
1.5    & -3         & 0.5220 & 0.5591        & 0.2575 & 1.760(-6)    & 2.919(-6)    & 9.985(-6)    \\
2.0    & -3         & 0.5859 & 0.6031        & 0.2490 & 1.580(-6)    & 3.284(-6)    & 9.763(-6)    \\
2.5    & -3         & 0.6916 & 0.6916        & 0.2486 & 1.497(-6)    & 3.727(-6)    & 9.415(-6)    \\
3.0    & -3         & 0.7563 & 0.7535        & 0.2484 & 1.342(-6)    & 4.335(-6)    & 8.929(-6)    \\
4.0    & -3         & 0.9569 & 0.8643        & 0.2654 & 1.181(-6)    & 5.255(-6)    & 8.094(-6)    \\
5.0    & -3         & 1.1936 & 0.9045        & 0.2941 & 1.091(-6)    & 5.892(-6)    & 7.485(-6)    \\
6.0    & -3         & 1.4495 & 0.9480        & 0.3153 & 1.036(-6)    & 6.307(-6)    & 7.088(-6)    \\
7.0    & -3         & 1.7101 & 0.9900        & 0.3313 & 1.091(-6)    & 6.608(-6)    & 6.796(-6)    \\
8.0    & -3         & 1.9968 & 1.1029        & 0.3397 & 1.374(-6)    & 6.780(-6)    & 6.630(-6)    \\
9.0    & -3         & 2.2105 & 2.2325 $^{b}$ & 0.2451 & 1.061(-6)    & 5.885(-6)    & 7.536(-6)    \\
\hline
0.8    & -4         & 0.5359 & 0.5594        & 0.2395 & 2.169(-7)    & 3.083(-7)    & 9.253(-7)    \\
1.0    & -4         & 0.5350 & 0.5733        & 0.2448 & 2.112(-7)    & 3.102(-7)    & 9.307(-7)    \\
1.2    & -4         & 0.5351 & 0.5697        & 0.2514 & 1.784(-7)    & 3.500(-7)    & 9.290(-7)    \\
1.5    & -4         & 0.5159 & 0.5500        & 0.2576 & 1.575(-7)    & 3.461(-7)    & 9.614(-7)    \\
2.0    & -4         & 0.6093 & 0.6323        & 0.2559 & 1.630(-7)    & 3.669(-7)    & 9.302(-7)    \\
3.0    & -4         & 0.7934 & 0.7855        & 0.2602 & 1.189(-7)    & 5.262(-7)    & 8.074(-7)    \\
4.0    & -4         & 0.9654 & 0.8769        & 0.2738 & 9.770(-8)    & 6.402(-7)    & 7.057(-7)    \\
5.0    & -4         & 1.1726 & 0.9030        & 0.2981 & 8.691(-8)    & 7.165(-7)    & 6.330(-7)    \\
6.0    & -4         & 1.4169 & 0.9463        & 0.3151 & 8.129(-8)    & 7.582(-7)    & 5.887(-7)    \\
7.0    & -4         & 1.6805 & 0.9943        & 0.3306 & 1.677(-7)    & 7.908(-7)    & 5.607(-7)    \\
8.0    & -4         & 1.9622 & 1.1158        & 0.3370 & 3.297(-7)    & 8.085(-7)    & 5.433(-7)    \\
\hline
0.8    & -5         & 0.5395 & -             & 0.2336 & 3.188(-8)    & 1.162(-8)    & 1.008(-7)    \\
1.0    & -5         & 0.5333 & -             & 0.2379 & 3.123(-8)    & 1.238(-8)    & 1.008(-7)    \\
1.1    & -5         & 0.5329 & 0.5725        & 0.2458 & 1.983(-8)    & 3.447(-8)    & 9.086(-8)    \\
1.2    & -5         & 0.5260 & 0.5631        & 0.2496 & 1.780(-8)    & 3.717(-8)    & 9.048(-8)    \\
1.5    & -5         & 0.5297 & 0.5766        & 0.2509 & 1.877(-8)    & 3.182(-8)    & 9.529(-8)    \\
2.0    & -5         & 0.6342 & 0.6520        & 0.2636 & 1.543(-8)    & 4.313(-8)    & 8.637(-8)    \\
3.0    & -5         & 0.8376 & 0.8071        & 0.2787 & 1.070(-8)    & 6.163(-8)    & 7.205(-8)    \\
4.0    & -5         & 1.0144 & 0.8668        & 0.3006 & 8.399(-9)    & 7.388(-8)    & 6.115(-8)    \\
5.0    & -5         & 1.1792 & 0.9048        & 0.3112 & 7.164(-9)    & 8.184(-8)    & 5.371(-8)    \\
6.0    & -5         & 1.3760 & 0.9328        & 0.3223 & 7.383(-9)    & 8.711(-8)    & 4.862(-8)    \\
7.0    & -5         & 1.6197 & 0.9866        & 0.3311 & 1.376(-7)    & 9.012(-8)    & 4.528(-8)    \\
8.0    & -5         & 1.6102 & 0.9828        & 0.3307 & 1.306(-7)    & 9.043(-8)    & 4.537(-8)    \\
9.0    & -5         & 2.1459 & 2.1889 $^{b}$ & 0.2527 & 6.090(-9)    & 8.852(-8)    & 4.753(-8)    \\
\hline
0.8    & $- \infty$ & 0.5083 & -             & 0.2330 & 0            & 0            & 0            \\
1.0    & $- \infty$ & 0.4993 & -             & 0.2333 & 0            & 0            & 0            \\
1.1    & $- \infty$ & 0.4901 & -             & 0.2342 & 0            & 0            & 0            \\
1.2    & $- \infty$ & 0.4966 & 0.5904        & 0.2393 & 0            & 0            & 0            \\
1.5    & $- \infty$ & 0.5737 & 0.6382        & 0.2496 & 0            & 0            & 0            \\
2.0    & $- \infty$ & 0.7244 & 0.6804        & 0.2900 & $< 10^{-15}$ & $< 10^{-15}$ & $< 10^{-15}$ \\
2.5    & $- \infty$ & 0.8727 & 0.7371        & 0.3186 & $< 10^{-15}$ & $< 10^{-15}$ & $< 10^{-15}$ \\
3.0    & $- \infty$ & 1.0001 & 0.7658        & 0.3390 & $< 10^{-15}$ & $< 10^{-15}$ & $< 10^{-15}$ \\
4.0    & $- \infty$ & 1.2154 & 0.8438        & 0.3536 & 9.608(-15)   & 1.072(-12)   & 1.786(-14)   \\
5.0    & $- \infty$ & 1.4228 & 0.9081        & 0.3674 & 1.156(-9)    & 9.119(-10)   & 4.001(-12)   \\
6.0    & $- \infty$ & 1.5135 & 0.9509        & 0.3687 & 9.794(-8)    & 6.753(-10)   & 2.123(-11)   \\
7.0    & $- \infty$ & 1.5374 & 1.0077        & 0.3708 & 2.304(-6)    & 9.401(-10)   & 2.386(-9)    \\
8.0    & $- \infty$ & 1.6830 & 1.1336        & 0.3618 & 1.165(-6)    & 1.288(-9)    & 7.135(-10)   \\
\hline
\end{tabular}

\medskip
$^{a}$ If the He-FDDM-R occurs, the values are taken just before the event. Otherwise, the values during the second dredge-up are given. See text for more detail.\\
$^{b}$ Carbon burning starts before the second dredge-up
\end{minipage}
\end{table*}

%% file: Table3.tex
\begin{table*}
 \centering
 \begin{minipage}{140mm}
  \caption{Characteristic values and chemical composition during the He-FDDM event}
  \label{tab:hefddm}
  \begin{tabular}{lllllllllll}
\hline
Mass & $\feoh$ & $M_{\textrm{core}}{}^{a}$ & $r_{\textrm{core}}$ & $T_{\textrm{core}}^{\textrm{p}}{}^{b}$ & $\log L_{\textrm{He}}^{\textrm{p}}{}^{b}$ & $Y$ & $X_{12}$ & $X_{14}$ & $X_{16}$ & Type \\
($\msun$) &   & ($\msun$) & ($10^{-2} R_{\sun}$) & ($10^{8}$ K) & ($\lsun$) &  &  &  &  & (R or A) \\
\hline
0.8  & -3 & 0.5043 & 1.40 & 2.336 & 6.968 & 0.2635 & 1.137e-2 & 5.578e-4 & 5.210e-4 & A \\ 
1.0  & -3 & 0.5112 & 1.43 & 2.292 & 6.703 & 0.2576 & 6.386e-3 & 3.797e-4 & 3.184e-4 & A \\ 
1.2  & -3 & 0.5130 & 1.47 & 2.234 & 6.523 & 0.2621 & 3.699e-3 & 2.439e-4 & 1.353e-4 & A \\ 
1.5  & -3 & 0.4976 & 1.32 & 2.436 & 7.563 & 0.2668 & 2.517e-3 & 2.576e-4 & 1.198e-4 & A \\ 
2.0  & -3 & 0.5771 & 1.37 & 2.513 & 7.213 & 0.2527 & 1.364e-3 & 1.077e-4 & 6.226e-5 & A \\ 
2.5  & -3 & 0.6740 & 1.32 & 2.692 & 7.114 & 0.2490 & 5.706e-4 & 4.802e-5 & 4.536e-5 & A \\ 
%3.0  & -3 & 2.03E+08 & 17713  & 2.36  & 0.7535 &  &  &  &  \\ 
\hline
0.8  & -4 & 0.5006 & 1.54 & 2.10  & 6.015 & 0.2691 & 6.124e-3 & 7.703e-4 & 2.433e-4 & A \\ 
1.0  & -4 & 0.5020 & 1.59 & 2.04  & 5.726 & 0.2659 & 2.935e-3 & 4.817e-4 & 8.496e-5 & A \\ 
1.2  & -4 & 0.5136 & 1.66 & 2.38  & 6.876 & 0.2611 & 3.783e-3 & 1.787e-4 & 2.384e-4 & A \\ 
1.5  & -4 & 0.4840 & 1.59 & 2.15  & 6.595 & 0.2666 & 2.824e-3 & 2.688e-4 & 2.101e-4 & A \\ 
2.0  & -4 & 0.6019 & 1.35 & 2.57  & 7.282 & 0.2591 & 9.993e-4 & 7.880e-5 & 4.794e-5 & A \\ 
%\hline
%3.0  & -4.8 & 1.77E+08 & 19953  & 2.41  & 0.8048 \\ 
%3.5  & -4.8 & 1.22E+08 & 21893  & 2.61  & 0.8495 \\ 
\hline
0.8  & -5 & 0.5395 & 2.14 & 2.58 & 10.859 & 0.4568 & 1.594e-2 & 3.110e-3 & 3.656e-6 & R \\ 
1.0  & -5 & 0.5333 & 3.07 & 2.52 & 10.830 & 0.3723 & 2.515e-2 & 1.685e-3 & 6.093e-6 & R \\ 
1.1  & -5 & 0.4999 & 1.35 & 2.41 & 7.340 & 0.2669 & 4.475e-3 & 4.834e-4 & 2.996e-4 & A \\ 
1.2  & -5 & 0.5087 & 1.29 & 2.57 & 7.792 & 0.2630 & 5.318e-3 & 3.131e-4 & 3.872e-4 & A \\ 
1.5  & -5 & 0.5386 & 1.24 & 2.97 & 8.919 & 0.2668 & 4.459e-3 & 1.457e-4 & 5.952e-4 & A \\ 
2.0  & -5 & 0.6222 & 1.29 & 2.73 & 7.580 & 0.2725 & 9.364e-4 & 8.871e-5 & 3.644e-5 & A \\ 
\hline
0.8  & $- \infty$ & 0.5083 & 1.94 & 2.41 & 9.954 & 0.3662 & 4.482e-3 & 2.635e-3 & 1.719e-6 & R \\ 
1.0  & $- \infty$ & 0.4993 & 1.98 & 2.33 & 9.840 & 0.3222 & 3.124e-3 & 1.515e-3 & 1.766e-6 & R \\ 
1.1  & $- \infty$ & 0.4901 & 2.04 & 2.24 & 9.719 & 0.3398 & 3.329e-3 & 1.863e-3 & 1.162e-6 & R \\ 
1.2  & $- \infty$ & 0.4053 & 1.69 & 1.76 & 5.012 & 0.2884 & 3.347e-5 & 4.271e-4 & 2.202e-5 & A \\ 
1.5  & $- \infty$ & 0.5008 & 1.37 & 2.38 & 7.168 & 0.2977 & 3.355e-3 & 2.409e-4 & 1.844e-4 & A \\ 
2.0  & $- \infty$ & 0.6948 & 1.19 & 2.86 & 7.785 & 0.2921 & 1.386e-4 & 6.742e-5 & 9.618e-7 & A \\ 
%2.5  & $- \infty$ & 3.46E+08 & 4.47  & 2.48  & 0.7370 &  &  &  &  & A \\ 
3.0  & $- \infty$ & - & - & - & - & - & - & - & - & A \\ 
\hline
\end{tabular}

\medskip
$^{a}$ values at the centre of hydrogen (R) or helium (A) burning shell, depending on the type of He-FDDM \\
$^{b}$ superscript ``p'' means the maximum value reached during the helium shell flash
\end{minipage}
\end{table*}

%% file: Table4.tex
\begin{table*}
 \centering
 \begin{minipage}{140mm}
\caption{Comparisons of abundances obtained by He-FDDM with previous works}
\label{tab:compare}
\begin{tabular}{lllllll}
\hline
Reference & Mass & $\feoh$ & $X_{\textrm{C}}$ & $X_{\textrm{N}}$ & $X_{\textrm{O}}$ & Note \\
\hline
This work & 0.8  & -3 & 1.137E-02 & 5.578E-04 & 5.210E-04 & \\ 
\citet{Campbell2008b} & 0.85 & -3 & 3.929E-04 & 5.788E-06 & 9.332E-05 & \\ 

This work & 1.0  & -3 & 6.386E-03 & 3.797E-04 & 3.184E-04 & \\ 
\citet{Iwamoto2004} & 1.0  & -2.7 & 8.691E-03 & 9.920E-04 & 6.152E-04 & a \\
\citet{Campbell2008b} & 1.0 & -3 & 2.005E-03 & 5.942E-05 & 2.597E-04 & \\ 

This work & 1.5  & -3 & 2.517E-03 & 2.576E-04 & 1.198E-04 & \\ 
\citet{Iwamoto2004} & 1.5  & -2.7 & 2.888E-03 & 3.188E-04 & 1.754E-04 & a \\

This work & 2.0  & -3 & 1.364E-03 & 1.077E-04 & 6.226E-05 & \\ 
\citet{Iwamoto2004} & 2.0  & -2.7 & 9.840E-04 & 1.803E-04 & 6.953E-05 & a \\
\citet{Campbell2008b} & 2.0 & -3 & 4.709E-04 & 1.259E-02 & 1.976E-04 & b \\ 

This work & 2.5  & -3 & 5.706E-04 & 4.802E-05 & 4.536E-05 & \\ 
\citet{Iwamoto2004} & 2.5  & -2.7 & 2.228E-04 & 6.763E-06 & 2.067E-05 & a \\

This work & 3.0  & -3 & 1.342E-06 & 4.335E-06 & 8.929E-06 & \\
\citet{Iwamoto2004} & 3.0  & -2.7 & 4.944E-05 & 7.291E-06 & 1.712E-05 & a \\
\citet{Campbell2008b} & 3.0 & -3 & 8.762E-04 & 2.233E-02 & 3.480E-04 & b \\ 

This work & 0.8  & -4 & 6.124E-03 & 7.703E-04 & 2.433E-04 & \\ 
\citet{Fujimoto2000} & 0.8  & -4 & 5.0E-03 & 8.8E-04 & 1.5E-04 & \\ 
\citet{Picardi2004} & 0.8  & -4.3 & 1.60E-02 & 3.64E-03 & 2.80E-03 & \\ 
\citet{Campbell2008b} & 0.85 & -4 & 3.190E-05 & 1.074E-05 & 2.479E-05 & \\ 

This work & 1.0  & -4 & 2.935E-03 & 4.817E-04 & 8.496E-05 & \\ 
\citet{Campbell2008b} & 1.0 & -4 & 4.884E-03 & 2.064E-04 & 1.729E-03 & \\ 

This work & 2.0  & -4 & 9.993E-04 & 7.880E-05 & 4.794E-05 & \\ 
\citet{Campbell2008b} & 2.0 & -4 & 1.770E-04 & 4.405E-03 & 9.311E-05 & c \\ 

This work & 0.8  & -5 & 1.594E-02 & 3.110E-03 & 3.656E-06 & \\ 
\citet{Picardi2004} & 0.8  & -5.3 & 9.66E-03 & 4.61E-03 & 3.98E-03 & \\ 
\citet{Cassisi1996} & 0.8  & -8.3 & -- & -- & -- & $\sim 10^{-3} \msun$ of protons are ingested. \\ 
\citet{Campbell2008b} & 0.85 & -5.45 & 6.050E-03 & 9.957E-04 & 3.890E-03 & \\ 

This work & 1.0  & -5 & 2.515E-02 & 1.685E-03 & 6.093E-06 & \\ 
\citet{Campbell2008b} & 1.0 & -5.45 & 1.837E-03 & 8.076E-05 & 1.039E-04 & \\ 

This work & 2.0  & -5 & 9.364E-04 & 8.871E-05 & 3.644E-05 & \\ 
\citet{Campbell2008b} & 2.0 & -5.45 & 4.962E-04 & 1.341E-02 & 2.059E-04 & c \\ 

This work & 3.0  & -5 & 2.104E-05 & 4.467E-07 & 6.322E-07 & \\ 
\citet{Campbell2008b} & 3.0 & -5.45 & 6.349E-04 & 1.577E-02 & 2.231E-04 & c \\ 

This work & 0.8  & $- \infty$ & 4.482E-03 & 2.635E-03 & 1.719E-06 & \\ 
\citet{Picardi2004} & 0.8  & $- \infty$ & 6.85E-03 & 6.60E-03 & 2.52E-03 & \\ 
\citet{Schlattl2002} & 0.82 & $- \infty$ & 1.597E-02 & 9.226E-03 & 2.012E-07 & d, atomic diffusion is included. \\
\citet{Fujimoto2000} & 0.8  & $- \infty$ & 4.0E-03 & 4.6E-03 & 2.5E-06 & \\ 
\citet{Campbell2008b} & 0.85 & $- \infty$ & 2.598E-05 & 2.437E-04 & 5.034E-04 & \\ 

This work & 1.0  & $- \infty$ & 3.124E-03 & 1.515E-03 & 1.766E-06 & \\ 
\citet{Siess2002} & 1.0  & $- \infty$ & 7.473E-04 & 7.854E-05 & 2.231E-04 & e \\ 
\citet{Picardi2004} & 1.0  & $- \infty$ & 6.12E-03 & 4.72E-03 & 5.84E-04 & \\ 
\citet{Hollowell1990} & 1.0  & $- \infty$ & 1.4E-03 & 2.6E-03 & -- & \\ 
\citet{Campbell2008b} & 1.0 & $- \infty$ & 1.844E-03 & 3.919E-03 & 4.333E-03 & \\ 

This work & 1.5  & $- \infty$ & 3.355E-03 & 2.409E-04 & 1.844E-06 & \\ 
\citet{Siess2002} & 1.5  & $- \infty$ & 6.905E-03 & 1.922E-05 & 2.356E-03 & e \\ 

This work & 2.0  & $- \infty$ & 1.386E-04 & 6.742E-05 & 9.618E-07 & \\ 
\citet{Siess2002} & 2.0  & $- \infty$ & 3.376E-03 & 6.718E-06 & 8.515E-05 & e \\ 
\citet{Campbell2008b} & 2.0 & $- \infty$ & 1.309E-04 & 3.432E-03 & 4.885E-05 & c \\ 

This work & 4.0  & $- \infty$ & 2.348E-13 & 8.004E-12 & 6.234E-14 & \\ 
\citet{Siess2002} & 4.0  & $- \infty$ & 4.453E-06 & 1.321E-03 & 1.167E-03 & e \\ 
\citet{Chieffi2001} & 4.0  & $- \infty$ & 2.940E-03 & 1.615E-05 & 2.958E-04 & e \\ 

This work & 5.0  & $- \infty$ & 1.156E-09 & 9.119E-10 & 4.001E-12 & \\ 
\citet{Siess2002} & 5.0  & $- \infty$ & 6.138E-05 & 2.963E-04 & 2.317E-05 & e \\ 

This work & 7.0  & $- \infty$ & 2.302E-06 & 2.707E-09 & 2.391E-09 & \\ 
\citet{Siess2002} & 7.0  & $- \infty$ & 8.939E-06 & 7.838E-05 & 1.750E-06 & e \\ 
\citet{Chieffi2001} & 7.0  & $- \infty$ & 1.109E-04 & 2.361E-04 & 1.247E-05 & e \\ 
\hline
\end{tabular}

\medskip
$^{a}$ They extend convection to the next shell beyond the Schwarzschild boundary. \\
$^{b}$ The surface abundances are not polluted by He-FDDM events, but polluted by TDU and HBB. \\
$^{c}$ The surface abundances are polluted by He-FDDM, TDU, and HBB. \\
$^{d}$ \citet{Anders1989} is adopted for solar abundances. \\
$^{e}$ The surface abundances are polluted by carbon ingestion. \\
\end{minipage}
\end{table*}

%% file: mn.bbl
\begin{thebibliography}{63}
\expandafter\ifx\csname natexlab\endcsname\relax\def\natexlab#1{#1}\fi

\bibitem[{{Alexander} \& {Ferguson}(1994)}]{Alexander1994}
{Alexander} D.~R., {Ferguson} J.~W., 1994, ApJ, 437, 879

\bibitem[{{Anders} \& {Grevesse}(1989)}]{Anders1989}
{Anders} E., {Grevesse} N., 1989, Geochim. Cosmochim. Acta, 53, 197

\bibitem[{Angulo {et~al.}(1999)Angulo, Arnould, Rayet, Descouvemont, Baye,
  Leclercq-Willain, Coc, Barhoumi, Aguer, Rolfs, Kunz, Hammer, Mayer,
  Paradellis, Kossionides, Chronidou, Spyrou, Degl'Innocenti, Fiorentini,
  Ricci, Zavatarelli, Providencia, Wolters, Soares, Grama, Rahighi, Shotter, \&
  Rachti}]{Angulo1999}
Angulo C., Arnould M., Rayet M., Descouvemont P., Baye D., Leclercq-Willain C.,
  Coc A., Barhoumi S., Aguer P., Rolfs C., Kunz R., Hammer J.~W., Mayer A.,
  Paradellis T., Kossionides S., Chronidou C., Spyrou K., Degl'Innocenti S.,
  Fiorentini G., Ricci B., Zavatarelli S., Providencia C., Wolters H., Soares
  J., Grama C., Rahighi J., Shotter A., Rachti M.~L., 1999, Nucl. Phys. A, 3

\bibitem[{{Aoki} {et~al.}(2006){Aoki}, {Frebel}, {Christlieb}, {Norris},
  {Beers}, {Minezaki}, {Barklem}, {Honda}, {Takada-Hidai}, {Asplund}, {Ryan},
  {Tsangarides}, {Eriksson}, {Steinhauer}, {Deliyannis}, {Nomoto}, {Fujimoto},
  {Ando}, {Yoshii}, \& {Kajino}}]{Aoki2006b}
{Aoki} W., {Frebel} A., {Christlieb} N., {Norris} J.~E., {Beers} T.~C.,
  {Minezaki} T., {Barklem} P.~S., {Honda} S., {Takada-Hidai} M., {Asplund} M.,
  {Ryan} S.~G., {Tsangarides} S., {Eriksson} K., {Steinhauer} A., {Deliyannis}
  C.~P., {Nomoto} K., {Fujimoto} M.~Y., {Ando} H., {Yoshii} Y., {Kajino} T.,
  2006, ApJ, 639, 897

\bibitem[{Aoki {et~al.}(2002)Aoki, Norris, Ryan, Beers, \& Ando}]{Aoki2002a}
Aoki W., Norris J.~E., Ryan S.~G., Beers T.~C., Ando H., 2002, ApJ, 567, 1166

\bibitem[{{Beers} {et~al.}(2007){Beers}, {Almeida}, {Rossi}, {Wilhelm}, \&
  {Marsteller}}]{Beers2007b}
{Beers} T.~C., {Almeida} T., {Rossi} S., {Wilhelm} R., {Marsteller} B., 2007,
  ApJS, 168, 277

\bibitem[{{Beers} {et~al.}(2005){Beers}, {Christlieb}, {Norris}, {Bessell},
  {Wilhelm}, {Allende Prieto}, {Yanny}, {Rockosi}, {Newberg}, {Rossi}, \&
  {Lee}}]{Beers2005b}
{Beers} T.~C., {Christlieb} N., {Norris} J.~E., {Bessell} M.~S., {Wilhelm} R.,
  {Allende Prieto} C., {Yanny} B., {Rockosi} C., {Newberg} H.~J., {Rossi} S.,
  {Lee} Y.~S., 2005, in IAU Symposium, Vol. 228, From Lithium to Uranium:
  Elemental Tracers of Early Cosmic Evolution, {V.~Hill, P.~Fran{\c c}ois, \&
  F.~Primas}, ed., pp. 175--183

\bibitem[{{Bromm} \& {Larson}(2004)}]{Bromm2004}
{Bromm} V., {Larson} R.~B., 2004, A\&AR, 42, 79

\bibitem[{{Campbell} \& {Lattanzio}(2008)}]{Campbell2008b}
{Campbell} S.~W., {Lattanzio} J.~C., 2008, A\&A, 490, 769

\bibitem[{{Cassisi} \& {Castellani}(1993)}]{Cassisi1993}
{Cassisi} S., {Castellani} V., 1993, ApJS, 88, 509

\bibitem[{{Cassisi} {et~al.}(1996){Cassisi}, {Castellani}, \&
  {Tornambe}}]{Cassisi1996}
{Cassisi} S., {Castellani} V., {Tornambe} A., 1996, ApJ, 459, 298

\bibitem[{{Chieffi} {et~al.}(2001){Chieffi}, {Dom{\'{\i}}nguez}, {Limongi}, \&
  {Straniero}}]{Chieffi2001}
{Chieffi} A., {Dom{\'{\i}}nguez} I., {Limongi} M., {Straniero} O., 2001, ApJ,
  554, 1159

\bibitem[{Christlieb {et~al.}(2002)Christlieb, Bessell, Beers, Gustafsson,
  Korn, Barklem, Karlsson, Mizuno-Wiedner, \& Rossi}]{Christlieb2002}
Christlieb N., Bessell M.~S., Beers T.~C., Gustafsson B., Korn A., Barklem
  P.~S., Karlsson T., Mizuno-Wiedner M., Rossi S., 2002, Nat, 904

\bibitem[{{Christlieb} {et~al.}(2004){Christlieb}, {Gustafsson}, {Korn},
  {Barklem}, {Beers}, {Bessell}, {Karlsson}, \&
  {Mizuno-Wiedner}}]{Christlieb2004b}
{Christlieb} N., {Gustafsson} B., {Korn} A.~J., {Barklem} P.~S., {Beers} T.~C.,
  {Bessell} M.~S., {Karlsson} T., {Mizuno-Wiedner} M., 2004, ApJ, 603, 708

\bibitem[{Frebel {et~al.}(2005)Frebel, Aoki, Christlieb, Ando, Asplund,
  Barklem, Beers, Eriksson, Fechner, Fujimoto, Honda, Kajino, Minezaki, Nomoto,
  Norris, Ryan, Takada-Hidai, Tsangarides, \& Yoshii}]{Frebel2005}
Frebel A., Aoki W., Christlieb N., Ando H., Asplund M., Barklem P.~S., Beers
  T.~C., Eriksson K., Fechner C., Fujimoto M.~Y., Honda S., Kajino T., Minezaki
  T., Nomoto K., Norris J.~E., Ryan S.~G., Takada-Hidai M., Tsangarides S.,
  Yoshii Y., 2005, Nat, 871

\bibitem[{Frebel {et~al.}(2006)Frebel, Christlieb, Norris, Aoki, \&
  Asplund}]{Frebel2006b}
Frebel A., Christlieb N., Norris J.~E., Aoki W., Asplund M., 2006, ApJ, L17

\bibitem[{Fujimoto {et~al.}(1990)Fujimoto, Iben~Jr., \&
  Hollowell}]{Fujimoto1990}
Fujimoto M.~Y., Iben~Jr. I., Hollowell D., 1990, ApJ, 580

\bibitem[{Fujimoto {et~al.}(2000)Fujimoto, Ikeda, \& Iben~Jr.}]{Fujimoto2000}
Fujimoto M.~Y., Ikeda Y., Iben~Jr. I., 2000, ApJ, L25

\bibitem[{Fujimoto {et~al.}(1995)Fujimoto, Sugiyama, Iben~Jr., \&
  Hollowell}]{Fujimoto1995}
Fujimoto M.~Y., Sugiyama K., Iben~Jr. I., Hollowell D., 1995, ApJ, 175

\bibitem[{{Garcia-Berro} \& {Iben}(1994)}]{GarciaBerro1994}
{Garcia-Berro} E., {Iben} I., 1994, ApJ, 434, 306

\bibitem[{{Gil-Pons} {et~al.}(2007){Gil-Pons}, {Guti{\'e}rrez}, \&
  {Garc{\'{\i}}a-Berro}}]{Gil-Pons2007}
{Gil-Pons} P., {Guti{\'e}rrez} J., {Garc{\'{\i}}a-Berro} E., 2007, A\&A, 464,
  667

\bibitem[{{Gil-Pons} {et~al.}(2005){Gil-Pons}, {Suda}, {Fujimoto}, \&
  {Garc{\'{\i}}a-Berro}}]{GilPons2005}
{Gil-Pons} P., {Suda} T., {Fujimoto} M.~Y., {Garc{\'{\i}}a-Berro} E., 2005,
  A\&A, 433, 1037

\bibitem[{Gilroy {et~al.}(1988)Gilroy, Sneden, Pilachowski, \&
  Cowan}]{Gilroy1988}
Gilroy K.~K., Sneden C., Pilachowski C.~A., Cowan J.~J., 1988, ApJ, 327, 298

\bibitem[{{Herwig}(2000)}]{Herwig2000}
{Herwig} F., 2000, A\&A, 360, 952

\bibitem[{{Herwig}(2003)}]{Herwig2003}
---, 2003, in Astronomical Society of the Pacific Conference Series, Vol. 304,
  Astronomical Society of the Pacific Conference Series, {Charbonnel} C.,
  {Schaerer} D., {Meynet} G., eds., pp. 318--323

\bibitem[{{Herwig} {et~al.}(2006){Herwig}, {Freytag}, {Hueckstaedt}, \&
  {Timmes}}]{Herwig2006}
{Herwig} F., {Freytag} B., {Hueckstaedt} R.~M., {Timmes} F.~X., 2006, ApJ,
  642, 1057

\bibitem[{Hollowell {et~al.}(1990)Hollowell, Iben~Jr., \&
  Fujimoto}]{Hollowell1990}
Hollowell D., Iben~Jr. I., Fujimoto M.~Y., 1990, ApJ, 245

\bibitem[{{Iben}(1975)}]{Iben1975}
{Iben} Jr. I., 1975, ApJ, 196, 525

\bibitem[{{Iben} {et~al.}(1992){Iben}, {Fujimoto}, \& {MacDonald}}]{Iben1992}
{Iben} I.~J., {Fujimoto} M.~Y., {MacDonald} J., 1992, ApJ, 388, 521

\bibitem[{{Iglesias} \& {Rogers}(1996)}]{Iglesias1996}
{Iglesias} C.~A., {Rogers} F.~J., 1996, ApJ, 464, 943

\bibitem[{{Itoh} {et~al.}(1996){Itoh}, {Hayashi}, {Nishikawa}, \&
  {Kohyama}}]{Itoh1996}
{Itoh} N., {Hayashi} H., {Nishikawa} A., {Kohyama} Y., 1996, ApJS, 102, 411

\bibitem[{{Itoh} {et~al.}(1983){Itoh}, {Mitake}, {Iyetomi}, \&
  {Ichimaru}}]{Itoh1983}
{Itoh} N., {Mitake} S., {Iyetomi} H., {Ichimaru} S., 1983, ApJ, 273, 774

\bibitem[{Iwamoto {et~al.}(2004)Iwamoto, Kajino, Mathews, Fujimoto, \&
  Aoki}]{Iwamoto2004}
Iwamoto N., Kajino T., Mathews G.~J., Fujimoto M.~Y., Aoki W., 2004, ApJ, 377

\bibitem[{{Komiya} {et~al.}(2009{\natexlab{a}}){Komiya}, {Habe}, {Suda}, \&
  {Fujimoto}}]{Komiya2009a}
{Komiya} Y., {Habe} A., {Suda} T., {Fujimoto} M.~Y., 2009{\natexlab{a}}, ApJ,
  696, L79

\bibitem[{{Komiya} {et~al.}(2009{\natexlab{b}}){Komiya}, {Suda}, \&
  {Fujimoto}}]{Komiya2009b}
{Komiya} Y., {Suda} T., {Fujimoto} M.~Y., 2009{\natexlab{b}}, ApJ, 694, 1577

\bibitem[{Komiya {et~al.}(2007)Komiya, Suda, Minaguchi, Shigeyama, Aoki, \&
  Fujimoto}]{Komiya2007}
Komiya Y., Suda T., Minaguchi H., Shigeyama T., Aoki W., Fujimoto M.~Y., 2007,
  ApJ, 367

\bibitem[{{Lattanzio}(1987)}]{Lattanzio1987}
{Lattanzio} J.~C., 1987, ApJ, 313, L15

\bibitem[{{Lau} {et~al.}(2008){Lau}, {Stancliffe}, \& {Tout}}]{Lau2008}
{Lau} H.~H.~B., {Stancliffe} R.~J., {Tout} C.~A., 2008, MNRAS, 385, 301

\bibitem[{{Lau} {et~al.}(2009){Lau}, {Stancliffe}, \& {Tout}}]{Lau2009}
---, 2009, MNRAS, 396, 1046

\bibitem[{{Lucatello} {et~al.}(2006){Lucatello}, {Beers}, {Christlieb},
  {Barklem}, {Rossi}, {Marsteller}, {Sivarani}, \& {Lee}}]{Lucatello2006}
{Lucatello} S., {Beers} T.~C., {Christlieb} N., {Barklem} P.~S., {Rossi} S.,
  {Marsteller} B., {Sivarani} T., {Lee} Y.~S., 2006, ApJ, 652, L37

\bibitem[{{Lucatello} {et~al.}(2005){Lucatello}, {Tsangarides}, {Beers},
  {Carretta}, {Gratton}, \& {Ryan}}]{Lucatello2005}
{Lucatello} S., {Tsangarides} S., {Beers} T.~C., {Carretta} E., {Gratton}
  R.~G., {Ryan} S.~G., 2005, ApJ, 625, 825

\bibitem[{{Masseron} {et~al.}(2009){Masseron}, {Johnson}, {Plez}, {Van Eck},
  {Primas}, {Goriely}, \& {Jorissen}}]{Masseron2009}
{Masseron} T., {Johnson} J.~A., {Plez} B., {Van Eck} S., {Primas} F., {Goriely}
  S., {Jorissen} A., 2009, ArXiv e-prints

\bibitem[{{McWilliam} {et~al.}(1995){McWilliam}, {Preston}, {Sneden}, \&
  {Searle}}]{McWilliam1995b}
{McWilliam} A., {Preston} G.~W., {Sneden} C., {Searle} L., 1995, AJ, 109, 2757

\bibitem[{{Miyaji} {et~al.}(1980){Miyaji}, {Nomoto}, {Yokoi}, \&
  {Sugimoto}}]{Miyaji1980}
{Miyaji} S., {Nomoto} K., {Yokoi} K., {Sugimoto} D., 1980, PASJ, 32, 303

\bibitem[{Nishimura {et~al.}(2009)Nishimura, Aikawa, Suda, \&
  Fujimoto}]{Nishimura2009}
Nishimura T., Aikawa M., Suda T., Fujimoto M.~Y., 2009, PASJ, accepted

\bibitem[{{Norris} {et~al.}(2007){Norris}, {Christlieb}, {Korn}, {Eriksson},
  {Bessell}, {Beers}, {Wisotzki}, \& {Reimers}}]{Norris2007}
{Norris} J.~E., {Christlieb} N., {Korn} A.~J., {Eriksson} K., {Bessell} M.~S.,
  {Beers} T.~C., {Wisotzki} L., {Reimers} D., 2007, ApJ, 670, 774

\bibitem[{{Norris} {et~al.}(1997){Norris}, {Ryan}, \& {Beers}}]{Norris1997}
{Norris} J.~E., {Ryan} S.~G., {Beers} T.~C., 1997, ApJ, 488, 350

\bibitem[{Picardi {et~al.}(2004)Picardi, Chieffi, Limongi, Pisanti, Miele,
  Mangano, \& Imbriani}]{Picardi2004}
Picardi I., Chieffi A., Limongi M., Pisanti O., Miele G., Mangano G., Imbriani
  G., 2004, ApJ, 609, 1035

\bibitem[{{Ritossa} {et~al.}(1996){Ritossa}, {Garcia-Berro}, \&
  {Iben}}]{Ritossa1996}
{Ritossa} C., {Garcia-Berro} E., {Iben} Jr. I., 1996, ApJ, 460, 489

\bibitem[{{Rossi} {et~al.}(1999){Rossi}, {Beers}, \& {Sneden}}]{Rossi1999}
{Rossi} S., {Beers} T.~C., {Sneden} C., 1999, in Astronomical Society of the
  Pacific Conference Series, Vol. 165, The Third Stromlo Symposium: The
  Galactic Halo, {Gibson} B.~K., {Axelrod} R.~S., {Putman} M.~E., eds., pp.
  264--268

\bibitem[{{Ryan} {et~al.}(1991){Ryan}, {Norris}, \& {Bessell}}]{Ryan1991}
{Ryan} S.~G., {Norris} J.~E., {Bessell} M.~S., 1991, AJ, 102, 303

\bibitem[{{Schlattl} {et~al.}(2002){Schlattl}, {Salaris}, {Cassisi}, \&
  {Weiss}}]{Schlattl2002}
{Schlattl} H., {Salaris} M., {Cassisi} S., {Weiss} A., 2002, A\&A, 395, 77

\bibitem[{{Siess} {et~al.}(2002){Siess}, {Livio}, \& {Lattanzio}}]{Siess2002}
{Siess} L., {Livio} M., {Lattanzio} J., 2002, ApJ, 570, 329

\bibitem[{{Spite} {et~al.}(2006){Spite}, {Cayrel}, {Hill}, {Spite}, {Fran{\c
  c}ois}, {Plez}, {Bonifacio}, {Molaro}, {Depagne}, {Andersen}, {Barbuy},
  {Beers}, {Nordstr{\"o}m}, \& {Primas}}]{Spite2006}
{Spite} M., {Cayrel} R., {Hill} V., {Spite} F., {Fran{\c c}ois} P., {Plez} B.,
  {Bonifacio} P., {Molaro} P., {Depagne} E., {Andersen} J., {Barbuy} B.,
  {Beers} T.~C., {Nordstr{\"o}m} B., {Primas} F., 2006, A\&A, 455, 291

\bibitem[{{Spite} {et~al.}(2005){Spite}, {Cayrel}, {Plez}, {Hill}, {Spite},
  {Depagne}, {Fran{\c c}ois}, {Bonifacio}, {Barbuy}, {Beers}, {Andersen},
  {Molaro}, {Nordstr{\"o}m}, \& {Primas}}]{Spite2005}
{Spite} M., {Cayrel} R., {Plez} B., {Hill} V., {Spite} F., {Depagne} E.,
  {Fran{\c c}ois} P., {Bonifacio} P., {Barbuy} B., {Beers} T., {Andersen} J.,
  {Molaro} P., {Nordstr{\"o}m} B., {Primas} F., 2005, A\&A, 430, 655

\bibitem[{Suda {et~al.}(2004)Suda, Aikawa, Machida, Fujimoto, \&
  Iben~Jr.}]{Suda2004}
Suda T., Aikawa M., Machida M.~N., Fujimoto M.~Y., Iben~Jr. I., 2004, ApJ, 476

\bibitem[{Suda \& Fujimoto(2006)}]{Suda2006}
Suda T., Fujimoto M.~Y., 2006, ApJ, 643, 897

\bibitem[{{Suda} {et~al.}(2007{\natexlab{a}}){Suda}, {Fujimoto}, \&
  {Itoh}}]{Suda2007}
{Suda} T., {Fujimoto} M.~Y., {Itoh} N., 2007{\natexlab{a}}, ApJ, 667, 1206

\bibitem[{Suda {et~al.}(2008)Suda, Katsuta, Yamada, Suwa, Ishizuka, Komiya,
  Sorai, Aikawa, \& Fujimoto}]{Suda2008}
Suda T., Katsuta Y., Yamada S., Suwa T., Ishizuka C., Komiya Y., Sorai K.,
  Aikawa M., Fujimoto M.~Y., 2008, PASJ, 60, 1159

\bibitem[{Suda {et~al.}(2006)Suda, Nishimura, Iwamoto, Aikawa, Fujimoto, \&
  Iben~Jr.}]{Suda2005}
Suda T., Nishimura T., Iwamoto N., Aikawa M., Fujimoto M.~Y., Iben~Jr. I.,
  2006, in Origin of Matter and Evolution of Galaxies 2005, Kubono S.,
  {et~al.}, eds., AIP Conference Proceedings, Melville, New York, pp. 59--64

\bibitem[{{Suda} {et~al.}(2007{\natexlab{b}}){Suda}, {Tsujimoto}, {Shigeyama},
  \& {Fujimoto}}]{Suda2007b}
{Suda} T., {Tsujimoto} T., {Shigeyama} T., {Fujimoto} M.~Y.,
  2007{\natexlab{b}}, ApJ, 671, L129

\bibitem[{Sugimoto \& Fujimoto(1978)}]{Sugimoto1978}
Sugimoto D., Fujimoto M.~Y., 1978, PASJ, 30, 467

\bibitem[{{Tsujimoto} \& {Shigeyama}(2006)}]{Tsujimoto2006}
{Tsujimoto} T., {Shigeyama} T., 2006, ApJ, 638, L109

\end{thebibliography}
